\documentclass[multphys,vecphys]{svmult}
\usepackage{amsxtra}
\usepackage{amstext}
\usepackage{amssymb}
\usepackage{latexsym}
\usepackage{epsfig}

\usepackage{graphicx}    
\usepackage{multicol}    

\newcommand{\be}{\begin{equation}}
\newcommand{\ee}{\end{equation}}
\newcommand{\bs }{\boldsymbol }
\newcommand{\rr}{{\boldsymbol r}}

\newcommand{\q}{{\boldsymbol q}}

\newcommand{\ep}{\epsilon}

\newcommand{\limite}[1]{ {{\raisebox{-.3cm}{$\textstyle\longrightarrow$}} \atop {\scriptstyle{#1}}}}

\begin{document}

\title*{Coherence and interactions in diffusive systems}
\author{Gilles Montambaux}
\institute{Laboratoire de Physique des Solides, Univ. Paris-Sud,
CNRS, UMR 8502, F-91405 Orsay Cedex, France.  --
\texttt{montambaux@lps.u-psud.fr} }
%
%
\maketitle

{\it Lectures notes for the "International School on Physics of
Zero and One Dimensional Nanoscopic Systems", Saha Institute of
Nuclear Physics, Calcutta, India, 1-9 February, 2006}
\bigskip

 This lecture is a tutorial
introduction to coherent effects in disordered electronic systems.
Avoiding technicalities as most as possible, I present some
personal points of view to describe well-known signatures of phase
coherence like weak localization correction or universal
conductance fluctuations.  I show how these physical properties of
phase coherent conductors can be simply related to the classical
return probability for a diffusive particle. The diffusion
equation is then solved in various appropriate geometries and in
the presence of a magnetic field. The important notion of quantum
crossing is developed, which is at the origin of the quantum
effects. The analogy with optics is exploited and the relation
between universal conductance fluctuations and speckle
fluctuations in optics is explained. The last part concerns the
effect of electron-electron interactions. Using the same simple
description, I derive qualitatively the expressions of the
Altshuler-Aronov anomaly of the density of states, and of the
correction to the conductivity. The last part, slightly more
technical, adresses the question of the lifetime of a
quasiparticle in a disordered metal.

\section{Introduction~: phase coherence and disorder}

Although the topic of this School mainly concerns nanoscopic
systems, this set of lectures is devoted to an intermediate range,
between the nanoscopic  and macroscopic scales, the so-called
mesoscopic regime \cite{AkkMon}. In this regime, the system to be
considered may be  large compared to the mean free path of the
electrons. Disorder plays then a very important role and, in the
so-called diffusive regime, the interplay between disorder and
quantum interference effects is crucial. This is the main subject
of these lectures. Here, electronic interactions will be treated
as a perturbation, in contrast with other topics discussed in this
 School where the electronic correlations may play the most important
role.
  I will try to
present some personal points of view in order to describe these
well-known signatures of phase coherence like weak  localization
or universal conductance fluctuations. The goal of these lectures
is to avoid technicalities as most as possible. The last part
concerns the effect of electron-electron interactions.

\medskip

To describe interference effects in electronics, it is useful to
compare with simple facts known in optics. The simplest experiment
with light is the two-slit Young experiment and phase coherent
effects considered in electronics are nothing but some more
sophisticated versions of the Young experiment. This two-slit
 experiment can be also performed with
electrons in vacuum \cite{webb} but here we shall consider
metallic wires, that are complex disordered media.

\begin{figure}[h]
\centerline{ \epsfxsize 3.5cm \epsffile{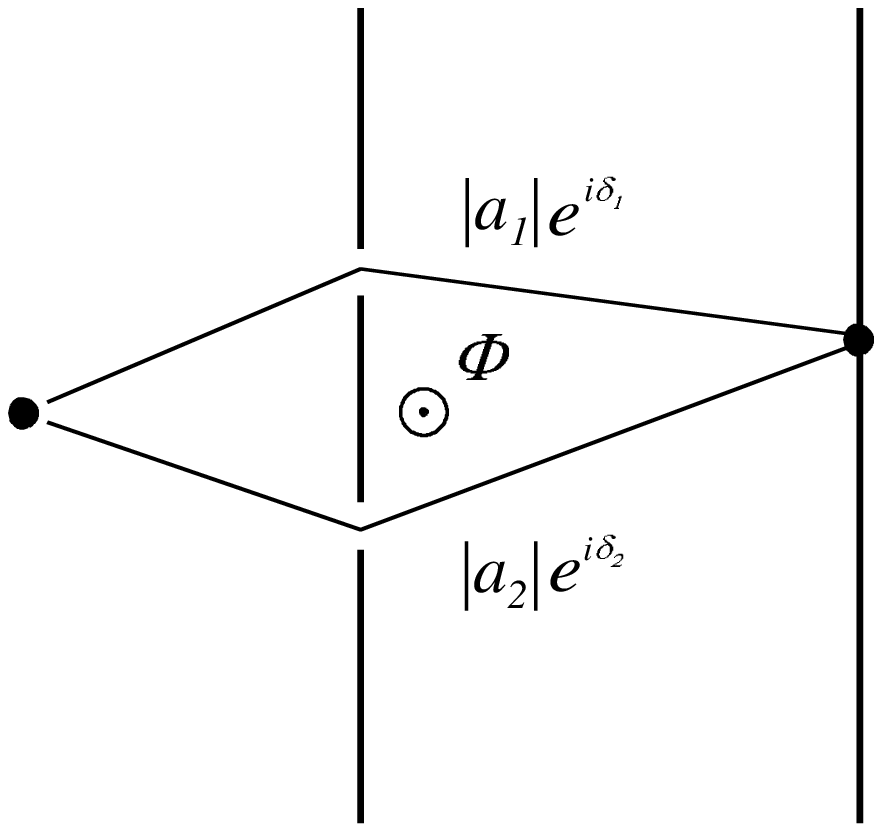} \vspace{.5cm}
\epsfxsize 3.5cm \epsffile{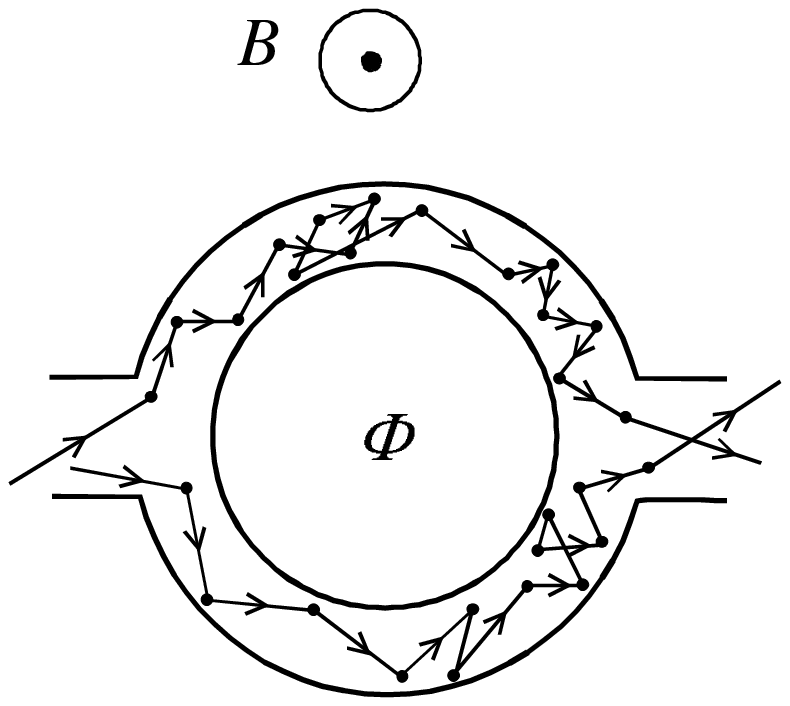} \vspace{.5cm}
\epsfxsize 3.5cm \epsffile{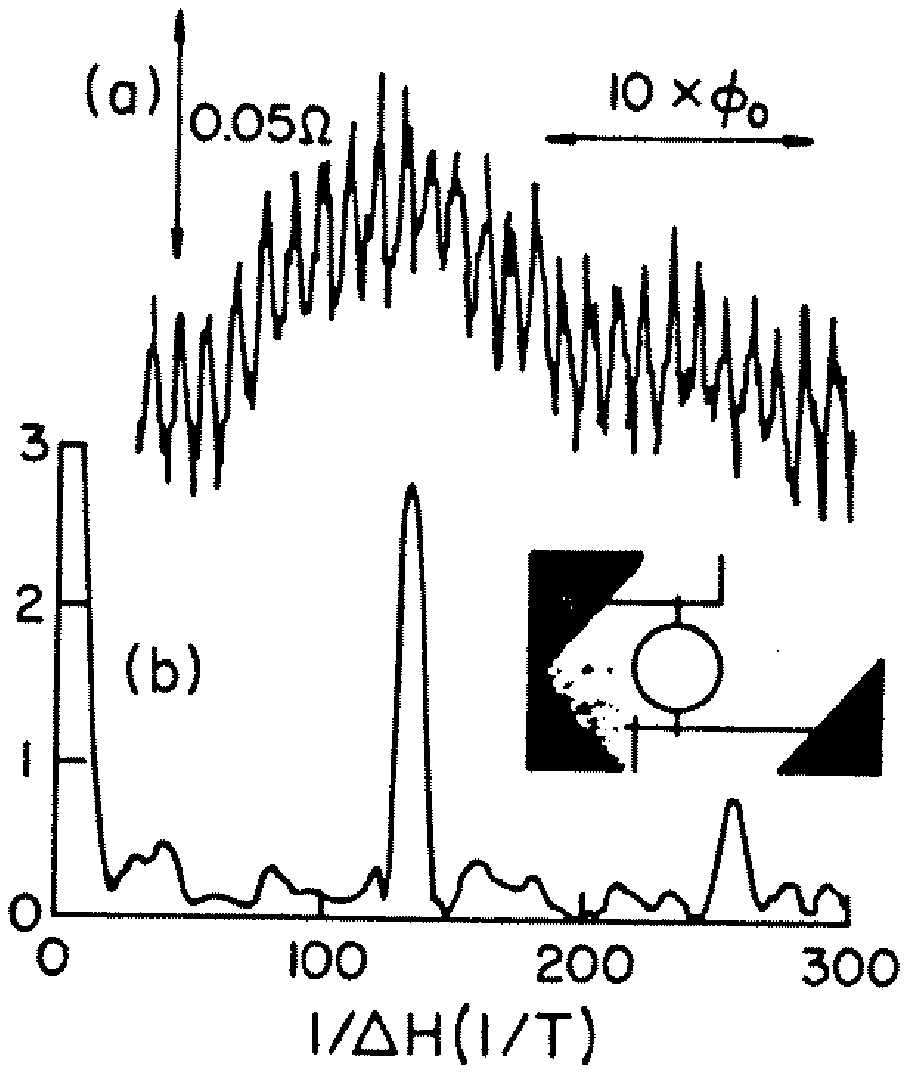} } \caption{\em Left~:
schematic representation of the Aharonov-Bohm effect. A flux tube
of flux $\phi$ is placed behind the two slits. Center~: Schematic
equivalent of the Aharonov-Bohm experiment in a metallic ring.
Right~: Magnetoresistance oscillations of a $Au$ ring and its
Fourier transform \cite{Webb851}. } \label{Introfig}
\end{figure}

In vacuum,  an electron beam is split in two parts and the
intensity is  measured on a screen. The topological equivalent in
a metal consists in a loop pierced by a magnetic flux and we
measure the current resulting from the interferences between the
two paths, see figure \ref{Introfig}. In optics the  way to probe
the interference pattern on the screen is to change the optical
path between the two trajectories, by changing the nature of the
medium, that is its optical index. For electrons,   the charge  is
coupled to the vector potential ${\bs A}$ and  these interference
pattern can be modified with a magnetic field.


If we try to transpose what is known from optics to electronics,
we measure a  current intensity which is proportional to the
probability for the electrons to traverse the loop. To calculate
this probability in quantum mechanics, we have to add the
contributions of two quantum amplitudes corresponding to the two
sides of the loop, and the current (a probability) is proportional
to the square of this quantum amplitude. Each quantum amplitude
$\psi_i$ has a phase $\varphi_i$~:

\be \psi_1=\psi \, e^{i \varphi_1} \qquad , \qquad \psi_2=\psi \,
e^{i \varphi_2} \ \ . \ee
 We have to sum the amplitudes and take the
modulus square. For the current, we get a classical term plus an
interference term~:

\be I \propto | \psi_1 + \psi_2|^2= 2 \psi^2 \, [ 1+ \cos(\phi_1 -
\phi_2)] \ \ . \label{2slits}\ee
 Classically, the current should be the sum of the two
currents, this corresponds to Ohm's law, but quantum mechanically
there is some additional phase effect. And if we apply a magnetic
field (this is the well-known Aharonov-Bohm effect
\cite{Aharonov59}),  there is an additional phase along each of
the two trajectories
\be \delta \varphi_1 = {e \over \hbar} \int_1  {\bs A} \cdot d
{\bs l} \qquad, \qquad \delta \varphi_2 = {e \over \hbar} \int_2
{\bs A} \cdot d {\bs l} \ \ ,  \ee
so that the phase difference is modulated by the circulation of
the vector potential along the closed loop formed by the two
trajectories,
\be \Delta \varphi= \delta \varphi_1 -\delta \varphi_2=  {e \over
\hbar} \oint  {\bs A} \cdot d {\bs l}= 2 \pi {\phi \over \phi_0} \
\ , \ee
and is proportional to  the  magnetic flux $\phi$  through the
loop. $\phi_0=h/e$ is the flux quantum.  This tells us that the
wave functions, the energy levels, thermodynamic and transport
properties must be function of this flux with period $\phi_0$. Of
course this does not tell us anything about the amplitude of these
oscillations. Unlike the case of the Young experiment with light
or for electrons in vacuum, the problem here is much more
complicated due to disorder. The question now is~: {\it do
interference effect survive in the presence of disorder?} Do
oscillations persist?

Indeed, some oscillations remain in the presence of disorder. The
pionner experiment founding  the field of mesoscopic physics was
performed by Webb {\it et al.} in 1985 \cite{Webb851}. They
measured the resistance of a ring of micronic size, in the
presence of a magnetic field (figure \ref{Introfig}). They found
that the resistance oscillates with the field, proving the
existence of an interference effect, even in the presence of
disorder. Two interesting features must be noted~: firstly, the
period of the oscillations is the flux quantum $\phi_0$ as
expected. Secondly, the typical amplitude of the oscillations of
the conductance is $\Delta G = \Delta R/R^2 \simeq e^2/h$.
\medskip

The necessary condition for these oscillations to exist is that
phase coherence is preserved, a condition which is obeyed in
vacuum. But in a metal, this phase coherence is broken because
electrons interact with other degrees of freedom ({\it e.g.}
phonons, other electrons, magnetic impurities).  Because of this
coupling, each electron can lose the memory of its phase. This
happens on a typical length, called the {\it phase coherence
length},  denoted by $L_\phi$, which depends on the coupling to
these degrees of freedom. Typically at 1K, it is of order of 1
$\mu$. This is the {\it mesoscopic range}. Of course, such
oscillations do not exist for a macroscopic ring, because phase
coherence is lost at the macroscopic level.

There is another important scale, due to static disorder, over
which electrons experience collisions over impurities. This scale
$l_e$ is named the {\it elastic mean free path}.  It is much
smaller than the phase coherence length. Since elastic collisions
do not break phase coherence, electrons may experience many
collisions without losing the memory of their phase. Each electron
keeps the memory of its phase typically on a scale $L_\phi \gg
l_e$. The physics we are going to discuss corresponds to length
scales which are much larger than the elastic mean free path
$l_e$, but smaller than $L_\phi$, so that the sample can be
considered as quantum mechanically coherent. This regime is called
the {\it mesoscopic regime}.
\medskip

We want to construct a theory for these oscillations. How to
explain their amplitude? How  can they survive disorder? Another
important key experiment  was performed by Sharvin and Sharvin on
a cylinder \cite{sharvin}, before the Webb's experiment on a
single ring. In this case, there are also oscillations, but
instead of being $\phi_0$-periodic, they have a period $\phi_0$/2.
The fundamental $\phi_0$ disappears and the experiment reveals the
second harmonics of these oscillations.

There is a simple way to understand this frequency doubling~:
assume that the cylinder can be viewed as a pile of independent
rings, it realizes an average of the oscillations of several
independent rings (similar effect for a network of rings
\cite{Pannetier84}). For a given ring, the oscillations have a
given phase which depends on the disorder configuration (This
phase is $0$ or $\pi$ for a two-terminal geometry
\cite{Buttiker887}). For another ring, the oscillations have a
different phase. When averaging over several rings (as is done in
a cylinder), because of this random phase, the oscillations
vanish.

 So, disorder is destructive for quantum interferences. However this simple argument would tell us that no oscillations should survive disorder averaging.
  This is not the case since the cylinder experiment shows oscillations with period $\phi_0/2$. In average, there are still oscillations, but with period $h/2e$.
This simple and very important fact tells us that {\it some
contributions  survive disorder averaging}. So the question is~:
how is it possible that some robust contribution survive disorder
averaging ?

\medskip

\begin{figure}
\begin{center}
\begin{tabular}{|c|}
  \hline
  \\

         \begin{minipage}{0.4\textwidth}
         \begin{center}
         Webb {\it et al.} experiment \medskip

         $\Downarrow$
         \medskip
         \end{center}

        Even in the presence of disorder, phase coherence is preserved on
        distances much larger than $l_e$ $\longrightarrow$
        sample specific interference effects, period $\phi_0$.

        \end{minipage}
\hfill \hspace{.5cm}

\begin{minipage}{0.4\textwidth}
\begin{center}
\vspace{-1. cm}

Sharvin-Sharvin experiment \medskip

$\Downarrow$ \medskip \end{center}

Some interference effects survive disorder averaging, period
$\phi_0/2$ $\longrightarrow$ pairing of trajectories.
\end{minipage} \\
 \\
  \hline
\end{tabular}
\end{center}
\caption{Conclusions that can be drawn from the two pionniering
experiments in mesoscopic physics.}
\end{figure}

\section{Important scales}

We shall consider weakly disordered metals, such that  the average
distance between two collision events is much larger than the
Fermi wave length~: $l_e \gg \lambda_F$. This condition allows for
a semiclassical description of electronic waves. Moreover we
assume that the typical size $L$ of the system is much larger than
the mean free path $l_e$ so that the electronic motion in the
sample is diffusive~:   electrons collide elastically many times
while traversing the system. Finally we assume that the system is
completely phase coherent, that is $L \ll L_\phi$. To summarize,
the system is weakly disordered,  diffusive and mesoscopic~:

\be \lambda_F \ll l_e \ll L \ll L_\phi  \ \ . \ee
We shall see that phase coherent effects may not disappear but are
simply reduced in the macroscopic limit $L_\phi < L$, so that we
shall also consider the case~:
 \be \lambda_F \ll l_e \ll L_\phi \ll L \ \ . \ee

\medskip

In strong disorder, when the mean free path becomes of order of
the Fermi wave length $k_F l_e \simeq 1$, interference effects are
strong and lead to localization of the electronic waves. This is
the domain where the electronic states are {\it exponentially
localized} in space, with   Anderson localization from extended to
localized waves. This topic will not be covered here. In the
opposite regime of very weak disorder, the mean free path becomes
so large (or the system is so small) that the mean free path
becomes larger that the system size. Collisions with impurities
are rare, and occur mainly on the boundaries of the system. This
the so-called {\it ballistic regime},  where the physics is mainly
driven by the structure of the boundaries, {\em i.e.} the shape of
the system. For most   shapes, the trajectories are chaotic. A
common method to describe this regime is the so-called Random
Matrix Theory  also used in other fields of physics like nuclear
physics. There are quite interesting common features between some
aspects of transport in chaotic dots and nuclear physics. Let us
also emphasize that the Random Matrix Theory of scattering or
transmission matrices can also be used to describe diffusive
systems \cite{Beenakker97}.
\medskip

Since the typical size $L$ of the system is  much larger than the
mean free path,  the electronic motion is diffusive.
 The average distance between collision events, the mean
free path is  related to the collision time $\tau_e$, $l_e=v_F
\tau_e$, $v_F$ being the Fermi velocity, since the electronic
motion between two collisions is ballistic. For times much longer
than the collision time $\tau_e$, the motion is diffusive and the
typical distance an electron can reach after a time $t$ scales
like
\be r^2 = D t \ \ , \label{rdet} \ee
where $D$ is the diffusion coefficient given by $D=v_F l_e/d$, $d$
being the space dimensionality. This relation tells us that for a
finite system of size $L$, a very important scale appears~: this
is the time for which an electron typically sees the boundaries of
the system. It is called the traversal time, or {\em Thouless
time}.  It is the time for  an electron to "realize" that the
system is finite. It is given by
\be \tau_D= {L^2 \over D} \ \ .  \label{Thoulesstime} \ee
To this characteristic time, is associated a characteristic
energy, the {\em Thouless energy} $E_c$~:

\be E_c = {\hbar \over \tau_D}={\hbar D \over L^2} \ \ . \ee This
energy scale plays a major role in the description of
thermodynamic and transport properties of mesoscopic diffusive
systems. For time scales smaller than $\tau_D$, the electron
propagates like in infinite space. The diffusive motion depends on
the space dimensionality of the system. On the other hand, in the
long time scale, the electronic motion explores the entire system,
this is the so-called ergodic regime.

Using eq.(\ref{rdet}), we can associate to the phase coherence
length $L_\phi$ a characteristic time, the phase coherence time
$\tau_\phi$~:

\be \tau_\phi ={L_\phi^2 \over D} \ \ . \ee This is the time
during which an electron keeps the memory of its phase.

\begin{figure}[h]
\centerline{ \epsfxsize 10cm \epsffile{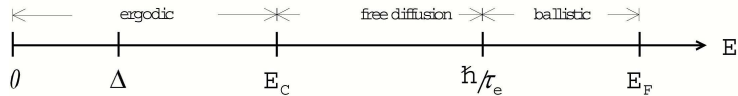} }
\caption{\em Characteristic  energy scales defining the different
regimes studied in coherent multiple scattering. Explanations are
given in the text.} \label{Echelles1}
\end{figure}
\medskip

Figure \ref{Echelles1} presents a scale of characteristic energies
(or inverse characteristic times). At short time scales, the
motion is ballistic. For times larger that $\tau_e$, the motion is
diffusive in free space. Above $\tau_D$, the motion is bounded,
this is the  ergodic regime. Then $\tau_\phi$ separates the
mesoscopic regime and the classical regime. If $\tau_\phi >
\tau_D$  ($L_\phi > L$),  the system is mesoscopic.
\medskip

In the diffusive regime, the space dimensionality $d$ plays an
important role.  Moreover, the one-dimensional case is somehow
special. In strictly one dimension, it is known that there no
diffusive regime since, in the presence of disorder, all states
are exponentially localized. So we shall not consider this case
but rather {\it quasi-}one-dimensional systems, with a transverse
width $a$, so that the real motion is three-dimensional (the
proper Schr\"odinger equation to be solved would be
three-dimensional with a quantization of the transverse component
of the wave vector), but the diffusion is one-dimensional. Instead
of having one transverse propagation channel, there are many
transverse channels (the transverse size is much larger than
$\lambda_F$). At short time (smaller than  a "transverse" Thouless
time $a^2/D$),
 diffusion is three-dimensional, but at larger times, there
is a one-dimensional propagation of the diffusion cloud. When
considering the transport through a wire ($L \gg a$), we shall be
interested in time scales necessary to traverse the wire, that is
times of order of $\tau_D$, which is much larger than the
transverse time, so that we can consider that at this time scale
the diffusion is one-dimensional.

\section{Classical probability and diffusion equation}
\label{CPDE}

The aim of  these lectures is to propose a qualitative description
of physical phenomena, trying to avoid sophisticated tools and
keeping in mind that we are essentially concerned by the
calculation of disordered averaged quantities.

An average quantity like the conductance basically  measures the
probability for electrons to cross the system. What is the nature
of this probability? Let us first spend some time  to describe the
probability $P(\rr,\rr')$ which describes the propagation of a
particle from a point $\rr$ to a point $\rr'$. In quantum
mechanics, this propagation is described by a probability {\it
amplitude}. This amplitude is called a Green's function
$G({\boldsymbol r},{\boldsymbol r}')$. We do not aim to develop
the theory of Green's functions. For our purpose here, it is
sufficient to note that there are many possible scattering
trajectories from ${\boldsymbol r}$ to ${\boldsymbol r}'$. Thus a
Green's function has the following structure~: it is the sum of
all the probability amplitudes corresponding to various multiple
scattering trajectories from ${\boldsymbol r}$ to ${\boldsymbol
r}'$, each trajectory being characterized by an amplitude and a
phase proportional to its action, that is its length
\cite{Feynmann}~:

\begin{equation} G({\boldsymbol r},{\boldsymbol r}') =
\sum_j A_j({\boldsymbol r},{\boldsymbol r}') \ \ .
\label{GArr'}\end{equation}
Now, we want to know the probability to find a particle at point
${\boldsymbol r}'$ if it has been injected at point ${\boldsymbol
r}$. The probability to go from ${\boldsymbol r}$ to ${\boldsymbol
r}'$ is the modulus square of the amplitude. From eq.
(\ref{GArr'}), we see that this probability is the sum of
amplitude squared terms, plus interference terms which pair
different trajectories $j$ and $j'$~:

 \begin{equation} |G({\boldsymbol r},{\boldsymbol r}')|^2 = \sum_{j,j'}
A_j({\boldsymbol r},{\boldsymbol r}')A_{j'}({\boldsymbol
r},{\boldsymbol r}')=\sum_j |A_j({\boldsymbol r},{\boldsymbol
r}')|^2 + \sum_{j' \neq j} A_j({\boldsymbol r},{\boldsymbol r}')
A_{j'}^*({\boldsymbol r},{\boldsymbol r}') \label{G2A2}
\end{equation}
which an obvious generalization of eq. (\ref{2slits}) for the
two-slit configuration. Since we know that, in quantum mechanics,
 one must add
amplitudes instead of intensities, the interference term (the
second term in eq. \ref{G2A2}) cannot be {\it a priori} neglected.
This second term describes interferences between {\it different}
trajectories $j$ and $j'$. Each  contribution in this sum has a
random phase which depends on the detail of the impurity
configuration.
 Since the phases are
uncorrelated, at the first level of approximation, we may expect
that the contribution of the interference term  cancels upon
disorder averaging. So quantum effects seem not to be  so
important because of the vanishing of this contribution. We shall
see however that this is not exactly the case. Within this
approximation, the second term cancels and the
 probability is essentially given by the sum of   intensities~:

\begin{equation} \overline{|G({\boldsymbol r},{\boldsymbol r}')|^2} = \overline{\sum_j
|A_j({\boldsymbol r},{\boldsymbol r}')|^2} \ \ .
\label{GRA3}\end{equation} We see that the phases have
disappeared. So the remaining term is completely classical.
Indeed, let us assume that some event changes the phase of the
amplitude $A_j$. The complex amplitude $A_j^*$ gets the opposite
phase, leaving the probability unchanged.

\begin{figure}[ht]
\centerline{ \epsfxsize 8cm \epsffile{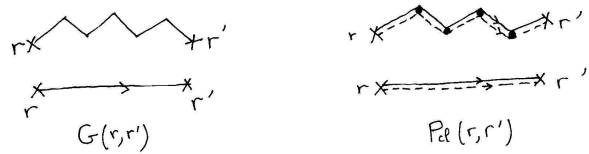}
 } \caption{\em  Schematic representations of a Green's function $G({\boldsymbol r},{\boldsymbol r}')$ and of the classical
  probability $P_{cl}({\boldsymbol r},{\boldsymbol r}')\propto  \overline{  \sum_j |A_j({\boldsymbol r},{\boldsymbol r}')|^2}$.
   The upper diagrams exhibit a few collision events, which are not represented on the lower diagrams.}
  \label{green}
\end{figure}

To have a simple picture of eq. [\ref{GRA3}], we represent on
figure \ref{green} a quantum amplitude as a line (it is rather a
sort of Brownian trajectory). Its complex conjugate is represented
as a dashed line. The first term in eq. (\ref{GRA3}) corresponds
to the pairing of a trajectory with its complex conjugate, and we
see immediately why the phase disappears. The quantity  $ \sum_j
\overline{|A_j({\boldsymbol r},{\boldsymbol r}')|^2}$ ressembles
the classical probability.  We call it a "{\it Diffuson}". To be
more precise, but without any proof, we define the
 probability $P({\boldsymbol r},{\boldsymbol r}',\omega)$ as

\begin{equation}
P({\bf r},{\bf r}',\omega) = {1  \over 2 \pi \rho_0}
 \overline{G_\epsilon({\bf r},{\bf r}')
 G^*_{\epsilon-\omega}({\bf r}',{\bf r})}
\label{proba}
\end{equation}
The Green's function and its complex conjugate are taken at
different energies (or frequencies) $\ep$ and $\ep - \omega$. One
can check that this probability is correctly normalized, that is
$\int_0^\infty P({\boldsymbol r},{\boldsymbol r}',t) d{\boldsymbol
r}'=1$, where $P({\boldsymbol r},{\boldsymbol r}',t)$ is the
Fourier transform of $P({\boldsymbol r},{\boldsymbol r}',\omega)$.
Starting from the Schr\"odinger equation in a random potential and
after disorder averaging, it is possible to show that in the limit
$k_F l_e \gg 1$,  the probability $P({\boldsymbol r},{\boldsymbol
r}',\omega)$ defined by (\ref{proba}) reduces to the Diffuson
$P_{cl}({\boldsymbol r},{\boldsymbol r}')\propto  \overline{
\sum_j |A_j({\boldsymbol r},{\boldsymbol r}')|^2}$. For slow
spatial variations,   $P_{cl}({\boldsymbol r},{\boldsymbol
r}',\omega)$ is the solution of a classical diffusion equation~:

\begin{equation}  \label{diff1}
\left( - i\omega - D \Delta \right)P_{cl}({\boldsymbol
r},{\boldsymbol r}',\omega)= \delta({\boldsymbol r}-{\boldsymbol
r}')
\end{equation}
where $D$ is  the diffusion coefficient. Doing this, we have only
considered classical contributions to the average (\ref{proba}).
We shall study later the corrections to this classical
probability.
\medskip

  Among the solutions of
this diffusion equation, one very important is the return
probability which enters in many physical quantities. It is the
probability $P(\rr,\rr,t)$ for an electron to return to its
original position after time $t$. I will also consider the space
integrated return probability~:
\begin{equation} P(t)=\int
P({\boldsymbol r},{\boldsymbol r},t) \, d{\boldsymbol r} \ \ .
\label{zdet}
\end{equation}
In free space, the solutions of  equation (\ref{diff1}) are simply
obtained from the Fourier transform

\begin{equation} \left({\partial \over \partial t} + D q^2 \right) P({\boldsymbol q},t) =
\delta (t)  \label{pdt1}\end{equation}
whose  solution $P(\q,t)$ is simply
\be P(\q,t)= e^{-D q^2 t} \ \ . \label{Pqt} \ee Fourier
transforming back, we find easily

\begin{equation}
\displaystyle P({\boldsymbol r},{\boldsymbol r}',t) =
\displaystyle {1 \over (4 \pi Dt)^{d/2}}e^{-|{\boldsymbol
r}-{\boldsymbol r}'|^2/4 D t} \ \ , \label{Pdiffinft}
\end{equation}
so that the return probability is given by
\be  P(\rr,\rr,t)= {1 \over (4 \pi D t)^{d/2}} \qquad \mbox{and}
\qquad  P(t)= {\Omega \over (4 \pi D t)^{d/2}} \ \ . \label{Pdetd}
\ee
where $\Omega$ is the volume of the system. The dependence on the
dimensionality $d$ of the return probability is crucial since it
will explain why dimensionality plays a so important role in
mesoscopic physics of diffusive systems.

\section{Conductance}

Now I wish to come to very simple and qualitative  considerations
about the conductance of a disordered system, which will be useful
 for these lectures.

\subsection{Classical conductance as the ratio of
two volumes}

Consider the conductance $G$. Since it has the dimensions of
$e^2/h$, we can introduce a dimensionless conductance $g$ as \be
g= G/(s e^2/h) \  \  . \ee Since this quantity is dimensionless,
it may be written  as the ratio of two physical quantities. For
example, by simple manipulations, it can be written as the ratio
of two energies~: $ g \propto  E_c /\Delta$, the Thouless energy
$E_c$  and the average level spacing $\Delta$. Here I would like
to write it as the {\it ratio of two volumes}. Let us start with
the classical Drude conductivity $\sigma_0$. From Einstein
relation, it is given by \be \sigma_0= s e^2 D \rho_0 \ \ ,
\label{sigmaeinstein} \ee
 where $D$ is the diffusion coefficient
and $\rho_0$ is the density of states at the Fermi level for one
spin direction. The factor $s=2$ accounts for spin degeneracy. By
Ohm's law, the conductance $G$ for a three-dimensional system is
given by $G=\sigma_0 S/L$, $S$ being the section and $L$ the
length of the sample. More generally for a hypercube of typical
size $L$ in $d$ dimensions, it is given by $G=\sigma_0 L^{d-2}$.
Introducing the Thouless time $\tau_D$ defined by
(\ref{Thoulesstime}),  let us rewrite the conductance as

\be G = s e^2 \rho_0 L^d /\tau_D \ \ . \label{Gthouless} \ee
The density of states at the Fermi level $\rho_0$ can be written
as $\rho_0= d A_d / 2\pi \lambda_F^{d-1} \hbar v_F$, where
$\lambda_F$ is the Fermi wavelength, $v_F$ is the Fermi velocity,
and $A_d$ is the volume of the unit sphere ($A_3= 4 \pi /3$,
$A_2=\pi$, $A_1=2$). An easy way to recover immediately this
result is to say that the total number of states is $ (k_F L)^d$,
so that by derivation with respect to the energy, we have
necessarily $\rho_0 \propto k_F^d/\ep_F \simeq k_F^{d-1}/ \hbar
v_F \propto 1/ \hbar v_F \lambda_F^{d-1}$. As a result, the
dimensionless conductance $g$ can be written as
 \be g = d A_d  \ {\Omega \over \lambda_F^{d-1} v_F \tau_D} \ \ , \label{gratio}\ee
where $\Omega=L^d$ is the volume of the system.  The dimensionless
conductance  quantity appears as the {\it ratio of two volumes},
the volume $\Omega$ of the system and the volume of a tube of
length $v_F \tau_D$ and of section $\lambda_F^{d-1}$. We shall see
later that this formulation will be quite useful to measure the
importance of interference effects.

\subsection{Conductance and transmission}

Our starting point to describe electric transport is the Landauer
formalism. Even staying  at a very qualitative level, this
formalism is quite natural since it expresses the {\it conductance
as a transmission coefficient} through the disordered sample.
\medskip

\begin{figure}[ht]
\centerline{ \epsfxsize 8cm \epsffile{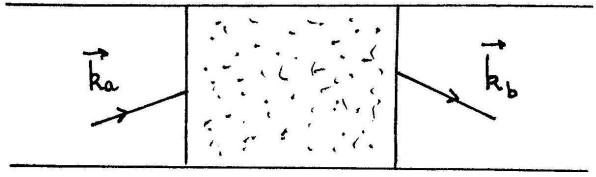} }
 \caption{\em  In the Landauer formalism, the conductance is related to the transmission coefficient between different incoming and outgoing channels.}
 \label{landauerfig}
\end{figure}

  Consider a disordered conductor of length $L$ and section $S = W^{d-1}$. It is connected to perfect conductors
  (figure \ref{landauerfig}) which can be considered as wave
guides where free electronic waves propagate. In this geometry,
the  transverse wave vectors of the eigenmodes (also called
channels) are quantized by transverse boundary conditions.  One
can define a transmission coefficient  $T_{ab}$ from an incoming
channel $a$ (ingoing wave vector ${\boldsymbol k}_a$) to an
outgoing channel $b$ (wave vector ${\boldsymbol k}_b$). The
Landauer formula reads~:
\begin{equation} G= s {e^2 \over h} \sum_{a,b} T_{ab} \ \ . \label{landauerm}\end{equation}
To calculate the number of transverse channels, one considers that
electrons are injected at the Fermi energy, {\it i.e.} such that
$|{\boldsymbol k}_a|=|{\boldsymbol k}_b|=k_{F}$. The transverse
component is quantized in units of $  2 \pi / W $. This
quantization imposes the number of channels. In $d=2$ and $d=3$,
their number is
\begin{equation} M_2 = {2 \pi k_{F} \over 2 \pi/ W} = { k_{F} W } \qquad \qquad
M_3 = {\pi k_{F}^{2} \over 4 \pi^{2}/ W^{2}} = { k_{F}^{2} S \over
4 \pi} \ \ .  \label{modestransv} \end{equation}
\medskip

\begin{figure}[ht]
\centerline{ \epsfxsize 7cm \epsffile{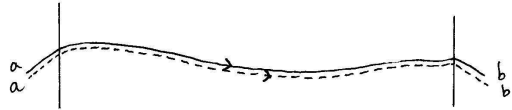} } \caption{\em
The conductance is proportional to the classical probability to
transmit  channel $a$ to  channel $b$ (summed over channels). The
object which represents this probability is the sum of
contributions of paired trajectories as introduced in figure
\ref{green}. We call it a "{\it Diffuson}".} \label{Diffusonab}
\end{figure}

Let us consider now the structure of the transmission coefficient
$T_{ab}$. It is the square of an amplitude  and it has, with minor
differences, the same structure as the probability $P({\boldsymbol
r},{\boldsymbol r}',\omega)$. The main difference is the
following~: instead of injecting a particle at a point
${\boldsymbol r}$ inside the sample, we inject a plane wave
${\boldsymbol k}_a$ from outside the sample. In particular, the
boundary conditions have to be treated properly. But, without
entering into  details, we may easily understand that, after
disorder averaging, the average transmission coefficient and
consequently the conductance can be related to the probability to
cross the sample. More precisely for a $3d$ sample, one can show
that the dimensionless conductance is \cite{AkkMon}

 \begin{equation}
g = {4 \over 9} M v_F P(0,L) \end{equation} where $P(0,L)$ is the
solution of the diffusion equation (\ref{diff1})  with appropriate
boundary conditions. It is given by $P(0,L)=l_e^2/D L$ so that
\begin{equation} g= {4 \over 3} M {l_e \over L} \ \ .  \end{equation} To obtain these results quantitatively, there
are some technicalities that we do not describe here
\cite{AkkMon}. What should be remembered is the message of fig.
\ref{Diffusonab}~: the conductance is proportional to the
classical probability to cross the sample. This statement is
sufficient to understand how coherence effects appear.

\section{Quantum corrections and quantum crossings}

\label{Qcqc}

The Diffuson is a classical object. It does not depend on the
phases of the complex amplitudes. In the diffusive regime, it is
solution of a diffusion equation. However, we may have to check
whether we have not left aside additional effects when throwing
out all the interference terms in relation (\ref{G2A2}). It turns
out that some of these terms   have quite interesting
consequences.

\begin{figure}[ht]
{\hfill \parbox[t]{6cm}{\epsfxsize \hsize \epsffile{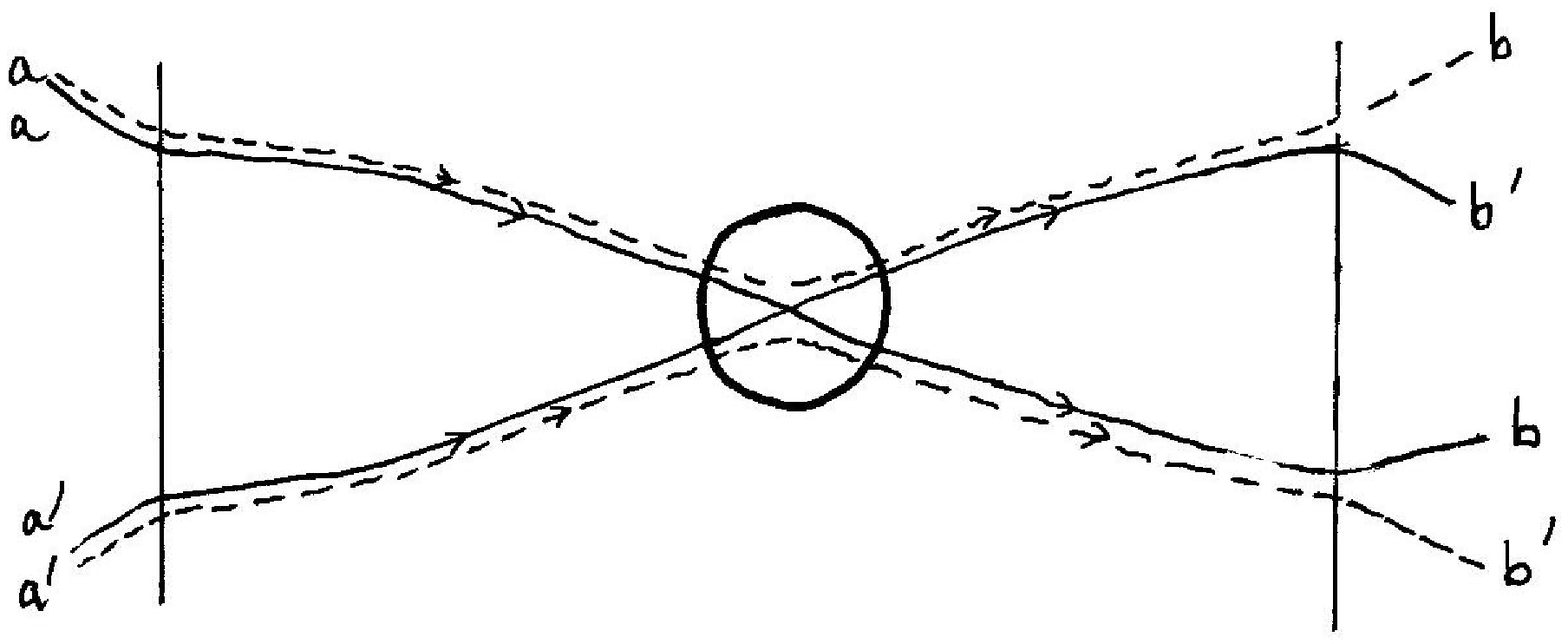}}
\hfill
\parbox[t]{4cm}{\vspace{-3cm} \epsfxsize \hsize
\epsffile{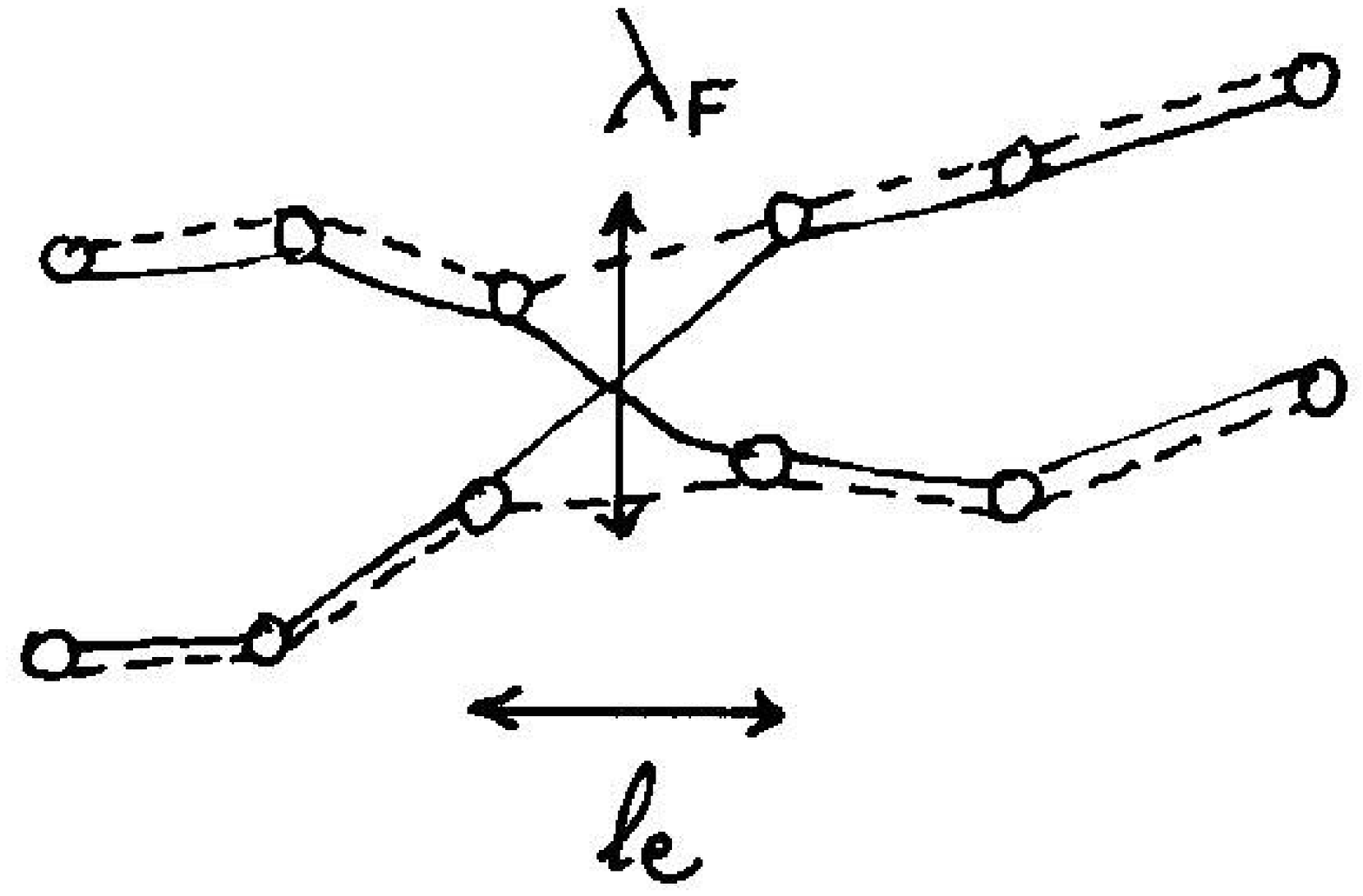}}
  \hfill }
  \caption{\em a) Crossing of two Diffusons. b) Detail~: the
volume of the intersection region is proportional to
$\lambda_F^{d-1} l_e$.} \label{Croisement}
\end{figure}

 Indeed, quantum effets can appear when two   Diffusons cross,
 or when a Diffuson crosses with itself. The notion of {\it quantum
 crossing}
 is extremely important because it is the source of quantum effects.
  The Diffuson being a classical object,
  coherence effects can only appear because of these quantum
 crossings. They are at the origin of the weak localization correction
 and of universal conductance fluctuations.
 Let us try to get some intuition about these crossing events.

Figure \ref{Croisement}.a shows that a crossing   mixes four
complex amplitudes which belong  to two incoming Diffusons  and
pair them differently. The two emerging Diffusons are built with
amplitudes $A_j$ and $A_{j'}$ coming respectively from each of the
incoming Diffusons. They have the same phase since they follow the
same path. The quantum crossing, also often called a Hikami box
in a more technical context, is thus an object whose role is to
permute the quantum amplitudes. It is necessarily short ranged,
because trajectories have to be as close as possible to each other
to avoid dephasing (figure \ref{Croisement}.b). Since it appears
between two successive collisions on impurities, and since the
phase mismatch between trajectories has to be smaller that $2
\pi$, one sees that the volume of this object is of order
$\lambda_F^{d-1} l_e$.

It is   important to evaluate the probability of occurrence of
such quantum crossings because it will be shown to be a measure of
the importance of quantum effects. Since the volume of a quantum
crossing is of order $\lambda_F^{d-1} l_e$, a Diffuson propagating
during a time $t$ can be seen  as
 an
effective object of length  ${\cal L} = v_F t $ and of "cross
section" $\lambda_F^{d-1}$. Thus, it has  a finite volume $v_F
\lambda_F^{d-1} t$. The probability $d p_\times(t)$ of crossing of
two Diffusons after a time $ d t$ in a volume $\Omega = L^{d}$ is
thus proportional to the ratio between the volume of a Diffuson
and the volume of the system~:

\begin{equation} d p_\times(t)= {\lambda_F^{d-1} v_F d t \over \Omega}\propto
{1 \over g} { d t \over \tau_D} \label{croix1}\end{equation}
where we have used (\ref{gratio}) to introduce the dimensionless
conductance $g$. Consider now an open system coupled to
reservoirs. The time needed to travel throughout the sample is the
Thouless time $\tau_D=L^2/D $. The probability of quantum crossing
during this time is given by

\begin{equation}  p_\times(\tau_D)=\int_0^{\tau_D} d p_\times(t)   =
{\lambda_F^{d-1} v_F \tau_D \over   \Omega }    \simeq {1 \over g}
\ \ .  \label{croix2}\end{equation}

This is exactly the inverse conductance ! I believe that this is
the most important message to understand phase coherence effects
in disordered systems. All these effects can be simply understood
in terms of quantum crossings and the probability of such
crossings which measures the importance of quantum mechanical
effects is simply given by the inverse of the dimensionless
conductance $g$.

In a good metal, the conductance $g$ is large, the volume of the
tube is small, electrons do not spend much time in the system and
quantum effects are very small. In the opposite limit, when $g$
becomes of order $1$, the volume of this tube is of the order of
the volume of the system. It is so big that the probability of
quantum crossing is of order $1$. This corresponds to the Anderson
regime where electronic waves are localized by strong disorder.
Here we shall not consider this regime but only the small disorder
regime where quantum effects remain small. The approach to
Anderson localization can be viewed as the proliferation of
quantum crossings.
\medskip

As a first qualitative but important conclusion of our discussion,
we see that classical transport is described by a conductance $G=s
g e^2/h$. Quantum corrections are smaller than classical terms by
a ratio $1/g$. This tells us immediately  that the quantum
corrections are of order $G/g$, that is $e^2/h$ !

\section{Weak localization}

\subsection{Weak localization and quantum crossings}

\begin{figure}[ht]
\centerline{ \epsfxsize 7cm \epsffile{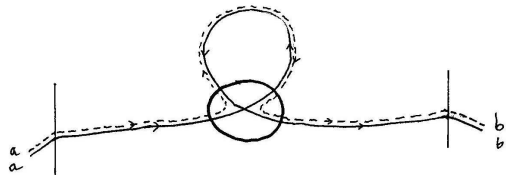}
 } \caption{\em Trajectory with a quantum crossing and a loop. In the loop, the two propagations are time reversed.}
  \label{loop}
\end{figure}

We have seen that the classical probability and the conductance
can be expressed as a sum
 of
contributions of pairs of complex conjugated trajectories.  Since
trajectories can have quantum crossings, they can form closed
loops (figure \ref{loop}). It turns out that in such a loop (whose
contribution is not included in eq. \ref{GRA3}),  the trajectories
are {\it time-reversed}.  One trajectory $j$ and its time reversed
$j^T$ go in opposite directions. If there is time-reversal
symmetry, they have the same action and thus they have exactly the
same phase. This phase can be quite complicated because it depends
on the disorder configuration but it is the same for both
trajectories. So the contribution of these loops does not cancel
on average. If the end points are far away like in fig.
\ref{loop}, the contribution of these new trajectories is small,
of order $1/g$, but it leads to an experimentally observable
effect~: the weak localization correction to the conductance. This
is a phase coherent effect because only trajectories of size
smaller than the phase coherence length $L_\phi$  contribute to
this additional contribution.

  At this point, I want to
stress that many presentations of weak localization correction
emphasize the  existence of a loop of opposite trajectories, but
do not insist on  the structure of the quantum crossing. This is
rather quite important, because this is what explains the
amplitude $1/g$ of the correction. This is where  phase coherence
is lost.
\medskip

Using the same type of argument as in the previous section, let us
evaluate the probability to have a loop for a trajectory which
travels through the  sample. Since there is a quantum crossing,
the probability is small, of order $1/g$. Moreover, it depends on
the distribution of loops in the disordered system. Let us call it
$P_{int}(t)$. For the probability of traversing the sample with a
loop,  we have~:

\begin{equation}  p_o(\tau_D)=\int_0^{\tau_D}P_{int}(t)\  d p_\times(t)  =   {1
\over g} \int_0^{\tau_D} P_{int}(t)  {d t\over \tau_D} \ \ .
\label{croix3}\end{equation}
We have also to remember that, because of decoherence in the
loops, only those with time $t$ smaller than $\tau_\phi$
contribute. The resulting probability to have trajectories with
loops of time smaller than $\tau_\phi$ is~:

\begin{equation}  p_o(\tau_\phi)=\int_0^{min(\tau_D,\tau_\phi)}P_{int} (t) \ d
p_\times(t)=  {1 \over g} \int_0^{min(\tau_D,\tau_\phi)}
P_{int}(t)  {d t\over \tau_D} \label{croix4}\end{equation}
where $P_{int}(t)$ is the probability to have loops of time $t$.
This leads to a relative correction to the conductivity (or to the
conductance) given by

\begin{equation}  {\Delta \sigma \over \sigma_0}= {\Delta G \over G_0} = -
p_o(\tau_\phi) \label{croix5}\end{equation}
 The sign of the correction
is negative
 because the trajectories $j$ and $j^T$ have opposite
momenta. This quantum correction to the classical Drude
conductivity is  called the {\it weak localization} correction
 \cite{wl,Bergmann84,Chakravarty86}.   The phase coherence
is broken by the coupling of the electrons to other degrees of
freedom or due to electron-electron interactions. Such coherence
breakdown is temperature dependent and can be phenomenologically
described by a temperature dependent phase coherence length
$L_\phi(T)= \sqrt{D \tau_\phi(T)}$~: trajectories larger than
$L_\phi$ do not  contribute to the weak localization correction.

\medskip

As we have seen above, the amplitude of the correction is
proportional to $P_{int}(t)$, the distribution  of loops. This
quantity is  nothing but the {\it return probability} after a time
$t$. It is not exactly the classical return probability, which is
the product of an amplitude with its complex conjugate (Diffuson).
Here it corresponds to the product of an amplitude with the
complex conjugate {\em time-reversed} amplitude. This object is
named a "{\it Cooperon}".  For closed trajectories, and with
time-reversal symmetry, the two contributions, Cooperon and
Diffuson,  are equal.  The return probability is thus doubled due
to quantum coherence.

Eqs. (\ref{croix4}, \ref{croix5}) have a meaning only in the
diffusive regime for which $t > \tau_e$ (otherwise a loop cannot
be formed). The contribution of the return probability has thus to
be integrated between $\tau_e$, the smallest time for diffusion,
and the phase coherence time $\tau_\phi$. Replacing the bounds by
exponential cutoffs, the weak localization correction can be cast
in the form \cite{exact}~:
\begin{equation}
\displaystyle \Delta G =-  2 s {e^2 \over h} \int_0^\infty
P_{int}(t)  \ (e^{- t/{\tau_\phi}}-e^{-t/\tau_e}) { dt \over
\tau_D} \ \ .
 \label{WL} \end{equation}
In order to evaluate $\Delta G$ in various situations, we now
study the diffusion equation and its solutions.

\subsection{How to solve the diffusion equation}
 In order to perform practical calculations, we have to calculate
this distribution of closed trajectories, that is, the return
probability. We have to solve a diffusion equation for this
interference part. It looks very much like a classical diffusion
equation (\ref{diff1}), but there is an important difference. To
account for magnetic field effects, it should be replaced by (in
time representation)~:

\begin{equation} \label{diff2}
\left[  {\partial \over \partial t} - D({\bf \nabla} + {2 i e
{\boldsymbol A}\over \hbar c })^2\right]P({\boldsymbol
r},{\boldsymbol r}',t)= \delta({\boldsymbol r}-{\boldsymbol
r}')\delta(t) \ \ .
\end{equation}
 The effect of the field is
described by a covariant derivative (like in a Schr\"odinger
equation) with an effective charge $2 e$ to account for the
pairing of trajectories.

 \medskip

To solve this equation (\ref{diff2}), let us notice that it is a
Green's equation, whose solutions are

\begin{equation} P({\boldsymbol r},{\boldsymbol r}',t)=\theta(t) \sum_n \psi^*_n({\boldsymbol r}) \psi_n({\boldsymbol r}')e^{-
E_n t} \ \ , \label{Pdifft}\end{equation}
where $\theta(t)$ is the step function and
    $\{E_n,\psi_n\}$   are the eigenvalues and eigenfunctions of the
eigenvalue equation  associated   to  (\ref{diff2})~:
\begin{equation}
 - D (\nabla_{\boldsymbol r} + {2 i e {\bf A} \over \hbar c})^2
\psi_n({\boldsymbol r}) = E_n  \psi_n({\boldsymbol r}) \ \ .
\label{schrodinger}
\end{equation}
 From eq. (\ref{Pdifft}), we find that the
integrated probability $P(t)$ has the simple form~:
\begin{equation}
P(t)=\theta(t) \sum_{n} e^{- E_n t} \ \ .   \label{partition}
\end{equation}
This important result tells us that in order to evaluate the weak
localization correction in any geometry, we simply need the
eigenvalues of the diffusion equation in the corresponding
geometry. We consider now a few examples.

\subsection{Dimension dependence of the weak localization
correction}

Consider an infinite system, or with size $L \gg L_\phi$. For free
diffusion in infinite space, the eigenvalues $E_n$ of the
diffusion equation are $D q^2$ and the return probability $P(t)$
is given by (\ref{Pdetd}). Since $P(t)$ is dimension dependent,
  we see than this weak localization  correction depends dramatically
on the space dimensionality. Inserting (\ref{Pdetd}) in
(\ref{WL}), and writing $\tau_\phi=L_\phi^2/D$, we obtain the
well-known results~:

 \be
\displaystyle \Delta g  = \left\{
\begin{tabular}{lll}
$ - \displaystyle  { L_\phi \over L}$ &  & $quasi-1d$ \\
\\
$- \displaystyle { 1 \over \pi } \ln {L_\phi\over l_e}$ &  & $d=2$ \\
\\
$-  \displaystyle  {1 \over 2 \pi } \left( {L \over l_e}- {L \over L_\phi}\right)$ &  & $d=3$%
\end{tabular}
\right. \label{sigmad} \ee
Since $L_\phi(T)$ varies as a power-law with temperature, we
obtain in particular the famous $\ln T$ dependence of the weak
localization  correction in $2d$.

It show be noticed that these results are meaningful if the
correction stays smaller than the classical conductance (which can
be written in the general form $g=  A_d (k_F L /2 \pi)^{d-1}\, l_e
/L$). This defines a characteristic length $\xi$ given  by $\Delta
g (\xi) \simeq g$, for which the weak-disorder perturbative regime
breaks down. In $1d$ and $2d$, it is given by \be \xi_{1d}= 2 l_e
\qquad, \qquad \xi_{2d}= l_e e^{\pi k_F l_e/2} \ \ , \ee and for a
quasi-$1d$ system $\xi_{q1d} \simeq M l_e$ where $M$ is the number
of channels. $\xi$ is the localization length.  For a review on
the strong localization regime, see for example \cite{Kramer93}.

\subsection{Finite systems, boundary conditions}
\label{FSBC}

 In a mesoscopic system, the
 cutoff time in (\ref{croix4}) is provided by $\tau_D$. In other words, the cutoff length in eqs. (\ref{sigmad}) is now
 the size $L$ of the system instead of $L_\phi$. Therefore, from (\ref{sigmad}), we see that in quasi-$1d$, the weak localization correction
 is universal in the sense that it is a number, independent of disorder strength ($l_e$). In $2d$ and $3d$, the integral
 (\ref{WL}) diverges at small time and is cut off by $\tau_e$, so that the correction cannot be universal \cite{AkkMon}.

 In order to calculate quantitatively the weak localization correction and the return probability in a finite system,
one must  be careful  to account properly for  correct boundary
conditions.

If the system is closed, electrons stay  inside the system, so
that $ P(t) \limite{t \rightarrow \infty} 1$. The correct boundary
condition is that the probability current vanishes at the boundary
(Neumann condition). Therefore  $q= n \pi /L $ with
$n=0,1,2,3,\cdots$. For $t \rightarrow \infty$, the contribution
of the zero mode in eq. (\ref{partition}) gives correctly $P(t)
\rightarrow 1$.

If the sample is perfectly connected to leads, electrons can leave
the sample and $ P(t) \limite{t \rightarrow \infty} 0$. The
probability at the boundary has to vanish (Dirichlet boundary
condition), because if it goes in the leads it never comes back in
the same state. The zero mode in now excluded,  $q= n \pi /L$ with
$n=1,2,3,\cdots$. $P(t) \rightarrow 0 $ when $t \rightarrow$ since
the particle leaves the box at large time. Inserting the
expression (\ref{partition}) of the return probability with
$E_n=n^2 E_c$ into (\ref{WL}) gives immediately, in the limit
$L_\phi \rightarrow \infty$~:
 \be \Delta g = - 2  \sum_{n \neq 0}
{1 \over \pi^2 n^2}= -{ 1 \over 3} \ \ . \ee This result is proper
to the perfectly connected wire.

\subsection{Magnetic field effects}

\subsubsection{Ring or cylinder geometry~: Sharvin-Sharvin oscillations}

Consider first the geometry of a ring pierced by a Aharonov-Bohm
flux.  In the presence of the flux, each closed trajectory
accumulates an Aharonov-Bohm phase $2 \pi \phi/\phi_0$, where
$\phi$ is the flux through the ring. The time-reversed trajectory
accumulates an opposite phase $-2 \pi \phi/\phi_0$, so that the
relative phase shift between the two trajectories is $4 \pi
\phi/\phi_0$, The fact that this relative phase between the two
time-reversed trajectories is twice the phase enclosed by one
trajectory is the reason why average quantities oscillate with
period $\phi_0/2=h/2e$.

 We need to calculate the return probability in this
geometry. This can be done directly by solving (\ref{diff2}). Here
let us proceed by simple arguments. Remember that in a $1d$
infinite space, the probability to go from $\rr$ to $\rr'$ is
given by (\ref{Pdiffinft}) with $d=1$. The return probability is
obtained by writing $\rr=\rr'$. On a ring, this would be the
return probability without making a loop, $1/\sqrt{4 \pi D t}$.
The return probability after one loop of perimeter $L$ necessarily
contains a term $e^{-L^2/4 D t}$. The accumulated phase is $4 \pi
\phi/\phi_0$ so that the flux dependent contribution of
trajectories making one loop is modulated by $\cos 4 \pi
\phi/\phi_0$. Adding together the contributions of $m$ loops, we
get simply the Fourier expansion of the flux dependent return
probability~:
\begin{equation}
P_{int}(t, \phi)= {L  \over \sqrt{4 \pi D t}}\sum_{m = -
\infty}^{+ \infty} e^{- m^2 L^2 /4 D t } \cos 4 \pi m \phi/\phi_0
\ \ .  \label{zring}
\end{equation}
Each harmonics of this expansion represents the  return
probability after $m$ loops around the ring. Inserting this
expression into (\ref{WL}) and after time integration, we obtain
easily

\begin{equation}
\displaystyle \Delta G (\phi)  = - s{e^2 \over  h } {L_\phi \over
L} \left( 1 +  2 \sum_{m =1}^{+ \infty} e^{- m L/L_\phi} \, \cos 4
\pi m \phi/\phi_0 \right) \label{sigmaring}
\end{equation}
which can be resumed to obtain
\be \displaystyle \Delta G (\phi)  = - s{e^2 \over  h } {L_\phi
\over L}{\sinh L/L_\phi \over \cosh L/L_\phi - \cos 4 \pi
\phi/\phi_0} \ \ .  \ee
The harmonics decay exponentially with their order, since they
correspond to longer and longer diffusive trajectories.

For a cylinder, there is a possibility for the electrons to
diffuse along the $z$ axis of the cylinder, so that (\ref{zring})
is simply multiplied by $L_z / \sqrt{4 \pi D t}$. Inserting this
new probability into eq. (\ref{WL}), we obtain

\begin{equation}
\displaystyle \Delta G (\phi)  = - s{e^2 \over \pi h } {L \over
L_z} \left[ \ln {L_\phi \over l_e} + 2 \sum_{m =1}^{+ \infty} K_0(
m L/L_\phi) \, \cos 4 \pi m \phi/\phi_0 \right]
\label{sigmasharvin}
\end{equation}
where $K_0$ is a modified Bessel function \cite{Abramowitz2}. The
$m=0$ term is the usual $2d$ result (\ref{sigmad}). The
contributions of the harmonics decay as $e^{-L/L_\phi}$. These
oscillations, predicted by Altshuler, Aronov and Spivak, where
observed by Sharvin and Sharvin in 1981
\cite{sharvin,Altshuler81}.

\subsubsection{$2d$ gas in a magnetic field}

In the ring geometry, all  pairs of diffusive trajectories would
pick the same phase $4 \pi \phi/\phi_0$. In a uniform magnetic
field, small and large trajectories  accumulate different fluxes
$\phi({\cal A})= B {\cal A}$ depending on their area ${\cal A}$.
So the return probability is balanced by the  average $\langle
\cos 4 \pi \phi({\cal A}) /\phi_0 \rangle_{\cal A}$ on the
distribution of areas ${\cal A}$ formed by the time-reversed
diffusive
 trajectories.

Let us start with a qualitative evaluation. Short trajectories
accumulate a flux smaller than the flux quantum and their
contribution survives. Large trajectories accumulate flux larger
than $\phi_0$ and their contribution vanishes. When  the magnetic
field increases,  the contribution of smaller and smaller
trajectories is progressively suppressed. Trajectories smaller
than  some field dependent characteristic length $L_B$
corresponding to $B L_B^2 \simeq \phi_0$ will not contribute. To
this length $L_B$ corresponds a characteristic time $\tau_B=
L_B^2/D \simeq \phi_0/ BD $, so that we can expect
\be \langle \cos 4 \pi \phi /\phi_0 \rangle_{\cal A}  \simeq e^{-
t / \tau_B} \ \ . \label{cost} \ee
Trajectories which enclose more than one flux quantum do not
contribute to the return probability. Because of this new cutoff
time, we can expect a field dependence of the weak localization of
the form
\be \Delta g = -{1 \over \pi} \ln {\min(L_\phi, L_B) \over l_e}
 \label{approx}\ee
instead of (\ref{sigmad}).

\begin{figure}[h!]
\centerline{ \epsfxsize 6cm \epsffile{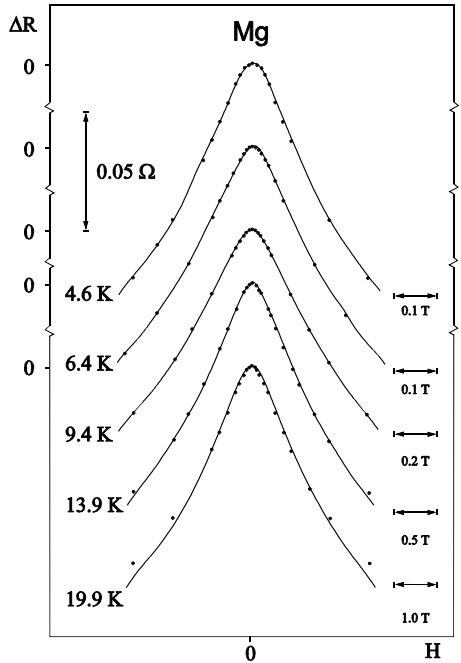} }
\caption{\em Magnetic field dependence of the magnetoresistance of
a $Mg$ film, for different temperatures. The points are
experimental results and the solid curves correspond to
(\ref{magnetcond2}). The time $\tau_\phi(T)$ is  a fitting
parameter \cite{Bergmann84}.} \label{Bergmann1}
\end{figure}

The exact calculation is straightforward starting from eq.
(\ref{partition}). The eigenvalues $E_n$ are solutions of an
effective Schr\"odinger for a free particle of mass $m= \hbar/2 D$
and charge $-2 e$ in a uniform field $B$. They are precisely the
Landau levels, namely

\begin{equation} E_n= (n+{1 \over 2}) {4 e D B \over \hbar} \ \ , \label{nivland} \end{equation}
where $n$ is an integer. The degeneracy of these levels for an
area $S$ is
 $g_n={2 e B \over h}S $. The  integrated return probability $P_{int}(t)$ is just given by
 the sum $\sum_n g_n e^{- E_n t}$, that is~:

\begin{equation}
P_{int}(t,B)= {B S/\phi_0 \over \sinh ( 4 \pi B D t / \phi_0)}\\
\label{ZcB}
\end{equation}
where  $\phi_0 = h/e$ is the flux quantum.
 This expression is nothing but the partition function of the
harmonic oscillator. In the limit $B \rightarrow 0$, one recovers
the result for  free diffusion~: $S / (4 \pi Dt)$. For large
times, $P_{int}(t,B)$ decreases exponentially with the
characteristic time $\tau_B = \phi_0 / 4 \pi B D$ introduced
qualitatively in (\ref{cost}). It describes the dephasing of time
reversed trajectories. Inserting eq. (\ref{ZcB})  in (\ref{WL}),
and performing the integral, we get~:

\begin{equation} \Delta g (B) = - {1   \over 2 \pi } \left[ \Psi \left({1
\over 2} + {\hbar \over 4 e D B \tau_e}\right) - \Psi \left({1
\over 2} + {\hbar \over 4 e D B \tau_\phi}\right) \right]
\label{magnetcond2} \end{equation}
where $\Psi$ is the digamma function. This expression corresponds
to the approximation (\ref{approx}). The weak localization
correction is negative and cancelled by the magnetic field. As a
result, a negative magnetoresistance is a well-known signature of
weak localization (figure \ref{Bergmann1}).  A magnetoresistance
measurement is a very interesting and useful tool to estimate
$\tau_\phi$. The correction cancels when $\tau_B \simeq
\tau_\phi$, that is for a field $B_\phi$ corresponding to a flux
quantum through an area $B_\phi L_\phi^2$. Doing the same
measurement at different temperatures is the usual method to
extract $\tau_\phi(T)$.

\section{Conductance fluctuations}

\subsection{Universality as a signature of quantum coherence}

\begin{figure}[h]
\begin{minipage}{0.4\textwidth}
\epsfxsize=\hsize \epsffile{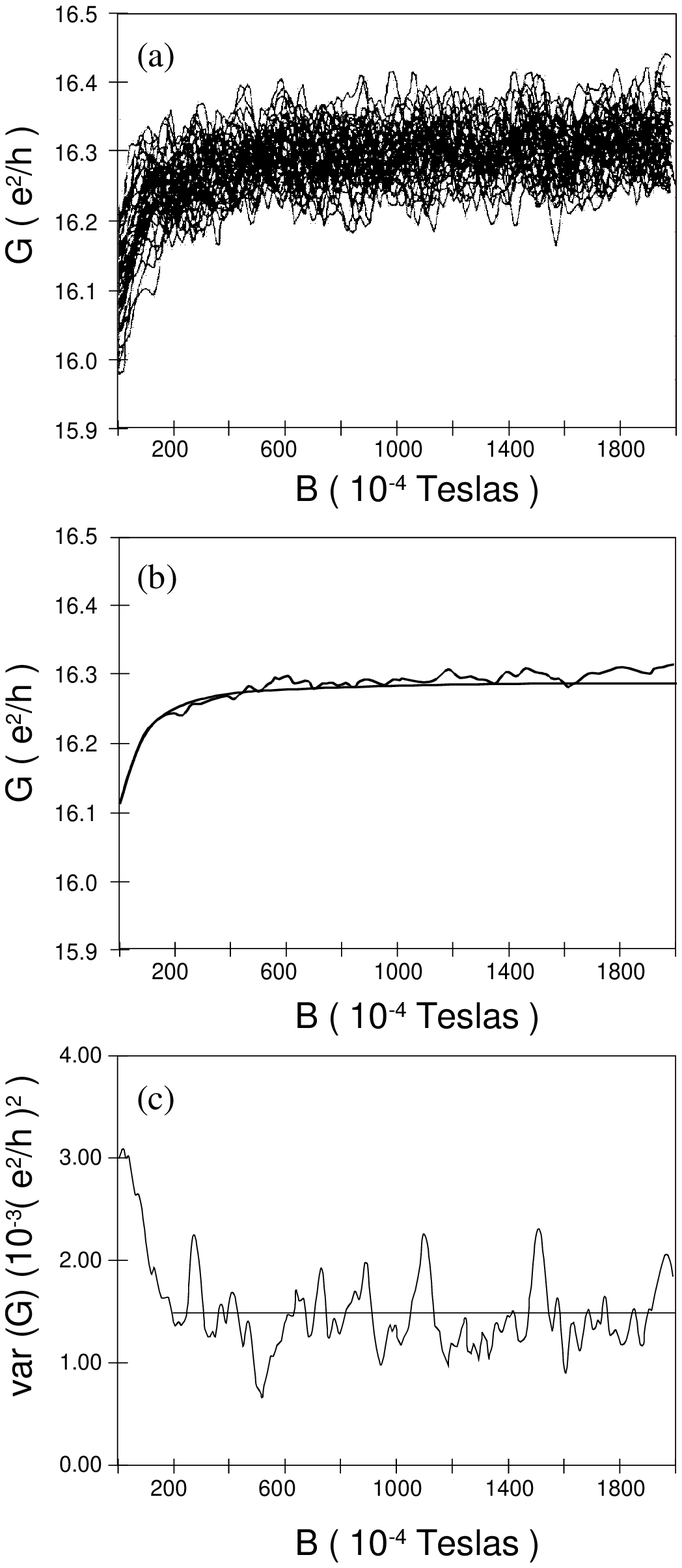}
\end{minipage}
\hfill
\parbox{.5\textwidth}{
\caption{\em Reproducible fluctuations of the magnetoconductance
in units of $e^2/h$, at  $T=45mK$ for  $Si$ doped $GaAs$. Fig. a
shows $46$ plots as function of the magnetic field, for the same
sample after successive annealing. Each plot corresponds to a
disorder configuration and is called a {\it magnetofingerprint}.
The amplitude of the fluctuations is smaller than $e^2/h$ because
$L > L_\phi$ (see eq. \ref{ucfred}). Fig. (b) presents the average
conductance versus field. The weak localization correction
disappears beyond a characteristic field. Above the same field,
the variance of the fluctuations displayed on fig. (c) is divided
by a factor $2$, corresponding to the destruction of the Cooperon
\protect\cite{Mailly92}.  } \label{mailly} }
\end{figure}

At a scale $L < L_\phi$, a conductor is a quantum object. Its
conductance depends on the interference pattern between all
diffusive trajectories. This interference pattern can be modulated
by external parameters, like a magnetic field or a gate voltage.
For example, figure \ref{mailly}.a represents the variation of the
conductance with the magnetic field, performed for $46$ different
samples (actually the same sample which has been annealed several
times, so that the impurity configuration has changed and the
interference pattern is different). It exhibits "fluctuations"
which are reproducible for a given configuration of disorder. They
are a "{\it fingerprint}" of this configuration. Figure
\ref{mailly}.b displays the average conductance, obtained by an
average over the 46 samples. One clearly sees the weak
localization correction, which is destroyed around some
characteristic field $B_\phi$. Interestingly, for the same
characteristic field, the variance of the fluctuations, displayed
on figure \ref{mailly}.c is reduced by a factor $2$.
\medskip

Universality of the conductance fluctuations is  a signature of
quantum transport. Classically, one would expect the system to be
considered  as an addition of large number of incoherent elements.
This number is of order ${\cal N}= (L/L_c)^d$, where $L_c$ would
be a correlation length, of the order of the mean free path. Then
one would expect relative fluctuations of the conductance $G$ of
order of

 \be {\delta G \over G} \simeq {1 \over \sqrt{{\cal N}}} =
\left({ L_c \over L}\right)^{d/2} \label{gLcL} \ee so that, since
$G$ varies as $L^{d-2}$~:

\be \delta G \propto L^{{d-4 \over 2}} \ \ ,  \ee and vanishes for
large $L$. The system is said to be self-averaging. But the fact
that the fluctuation $\delta G$ stays actually finite means that
there are strong correlations due to quantum coherence. Moreover,
if one considers fully coherent ($L < L_\phi$) systems with quite
different conductances, a good metal, a bad metal, or a
semiconductor, one finds that the amplitude of the "oscillations"
is always the same~: it does not depend on the disorder. It is
universal, of order $e^2/h$. {\it A priori}, we are not so
surprised that these fluctuations are universal since our simple
argument of section \ref{Qcqc}  showed that all quantum effects
have to
 be of order $e^2/h$.
 \medskip

\begin{figure}[h]
\centerline{ \epsfxsize 7cm \epsffile{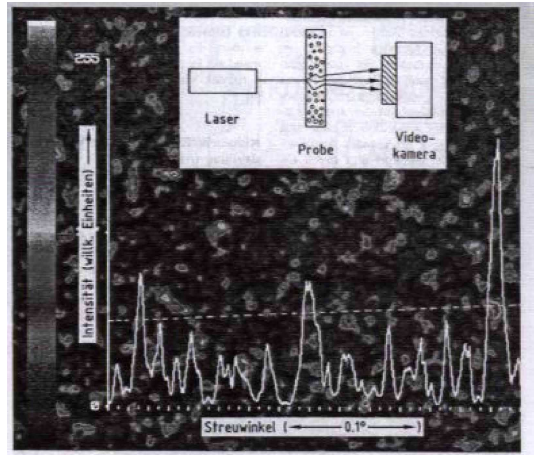} } \caption{\em A
typical speckle pattern. The white and noisy curve represents the
angular dependence of the light  intensity along the cut
represented by the dashed line. The relative  fluctuations are of
order unity  (courtesy of  G. Maret).} \label{Tav}
\end{figure}

\subsection{conductance fluctuations and speckle correlations
in optics}  Here it is quite useful to compare the physics of
electronic
 transport with similar physics in optics where one measures the
 fluctuations of a transmission coefficient. In optics the light
 scattered by a diffusing medium forms a {\it speckle} pattern on a screen, and
 we want to describe the speckle fluctuations (This is exactly a generalization of Young experiments.
 Two slits produce well defined fringes. Here the diffusing medium,
 like {\it e.g.} colloidal suspension, produces a complicated pattern called a speckle).
 A laser beam is sent on a diffusing medium along an incident
 direction $a$ and the diffused intensity is measured along a
 direction $b$. The speckle pattern displayed on figure \ref{Tav} represents the intensity measured along a direction $b$
  for a fixed
 incident direction $a$. So a given intensity on the screen
 represents the transmission coefficient $T_{ab}$ from an incident
 direction $a$ to an emergent direction $b$. We notice immediately that there are  black
spots, meaning that the relative  fluctuations of this coefficient
are of order $1$. This is the Rayleigh law~:
\be \overline{\delta T_{ab}^2}=
\overline{T}\hspace{-.1cm}_{ab}\hspace{.1cm}\hspace{-.2cm}^2\hspace{.2cm}
\ \ . \label{RL} \ee
In electronics however, relative fluctuations of the conductance
are very small, of order $1/g^2$. Here we want to understand why
fluctuations are large in optics and small in electronics, namely
to compare conductance fluctuations and fluctuations of
transmission coefficient.
\medskip

A very convenient way to link the two fields of optics and
electronics is to use the Landauer formalism, which explicitly
expresses the conductance as a transmission coefficient. This is
formalized by Landauer formula $g=\sum_{ab} T_{ab}$
(\ref{landauerm}).  The main difference between optics and
electronics is that in optics, it is possible to measure each
transmission coefficient $T_{ab}$ while in electronics the
conductance is related by eq. (\ref{landauerm}) to a {\it sum}
over all incoming and outgoing channels.

The average transmission  is a probability. It is a sum of
contributions of paired trajectories. Assuming that the angular
(or channel) dependence of $\overline{
T}\hspace{-.1cm}_{ab}\hspace{.1cm}$ is negligible, all the
channels contribute equally to the conductance, so that, from
(\ref{landauerm})~:

\be {\overline G}= {e^2 \over h} M^2 \, {\overline{
T}\hspace{-.1cm}_{ab}\hspace{.1cm}} \ee
and the average transmission coefficient is thus equal to \be
{\overline{ T}\hspace{-.1cm}_{ab}\hspace{.1cm}}= {g \over M^2}  \
\ . \ee

\begin{figure}[htb]
\centerline{ \epsfxsize 8cm \epsffile{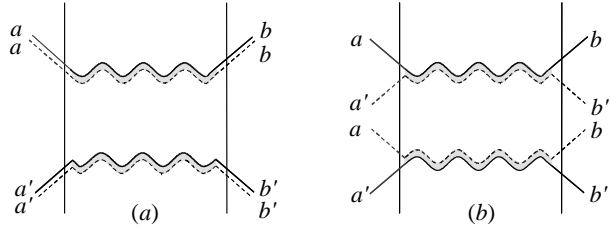} }
\caption{\em  Schematic representation of the two contributions to
the product $\overline{ T_{ab} T_{a'b'}}$. The first (a)
corresponds to the product $\overline{
T}\hspace{-.1cm}_{ab}\hspace{.1cm} \, \overline{
T}\hspace{-.1cm}_{a'b'}\hspace{.1cm}$. The second (b) gives a
contribution to the correlation function that we shall denote  by
$\overline{\delta T_{ab} \, \delta T_{a'b'}}^{C_1}$.}
\label{Ampfrauc1}
\end{figure}

Now we want to calculate the correlation between two transmission
coefficients, that is the function $\overline{ T_{ab} T_{a'b'}}$.
This quantity is the product of two average transmission
coefficients plus a correlation term~:

\be \overline{T_{ab} T_{a'b'}}= \overline{
T}\hspace{-.1cm}_{ab}\hspace{.1cm} \ \overline{
T}\hspace{-.1cm}_{a'b'}\hspace{.1cm}+ \overline{\delta T_{ab} \,
\delta T_{a'b'}} \ \ .  \ee
The correlation term is constructed by  pairing of trajectories
corresponding to different transmission coefficients (Figure
\ref{Ampfrauc1} ).

One sees that there is no dephasing between the diffusive paired
trajectories, expect outside the sample since $a$ and $a'$ ($b$
and $b'$) may correspond to different incoming (outgoing)
channels. The second term is therefore of the form $\overline{
T}\hspace{-.1cm}_{ab}\hspace{.1cm}\hspace{-.2cm}^2\hspace{.1cm}
f(a,a',b,b')$ where $f$ is a short range function which vanishes
rapidly as soon as $a \neq a'$ or $b \neq b'$. If $a$ and $b$ are
angular directions, $f$ is a rapidly decreasing function of the
angles. In a wave guide geometry, where the modes are quantized,
$f=\delta_{aa'} \delta_{bb'}$. This contribution to the
correlation function $\overline{\delta T_{ab} \, \delta T_{a'b'}}$
is called $C_1$~:

\be \overline{\delta T_{ab} \, \delta T_{a'b'}}^{C_1}=\overline{
T}\hspace{-.1cm}_{ab}\hspace{.1cm}\hspace{-.2cm}^2\hspace{.2cm} \,
\delta_{aa'} \delta_{bb'} \ \ . \label{C1} \ee
 For $a=a'$
and $b=b'$, we obtain the Rayleigh law (\ref{RL}), that is
\be {\overline T_{ab}^2 }= 2 \ \overline{
T}\hspace{-.1cm}_{ab}\hspace{.1cm}\hspace{-.2cm}^2\hspace{.2cm} \
\ . \ee
The amplitude of the fluctuations is of the order of the average.
This explains why there are black spots on figure \ref{Tav}.

In order to calculate the conductance fluctuations, we have to sum
over all incoming and outgoing channels~:
\be \overline{\delta g^2} =\sum_{aa'bb'} \overline{ \delta T_{ab}
\delta T_{a'b'}}= \sum_{aa'bb'}\overline{
T}\hspace{-.1cm}_{ab}\hspace{.1cm}\hspace{-.2cm}^2\hspace{.2cm}
\delta_{aa'}\delta_{bb'}= M^2 \,  \overline{
T}\hspace{-.1cm}_{ab}\hspace{.1cm}\hspace{-.2cm}^2\hspace{.2cm} =
{g^2 \over M^2} \ll 1  \ \ . \ee
The sum is small since most of the terms are negligible. So our
picture explains the important fluctuations of $T_{ab}$ but {\it
not} the amplitude of the conductance fluctuations. This means
that additional contributions to the correlation function
$\overline{\delta T_{ab} \, \delta T_{a'b'}}$ may have been
forgotten.

 A next contribution is obtained by pairing trajectories in a
different way. One possibility is to exchange the quantum
amplitudes, and to have one crossing as shown in figure
\ref{onetwocrossings}.a. This contribution is smaller by a factor
$1/g$ , so that its contribution to the correlation function
$\overline{ \delta T_{ab} \delta T_{a'b'}}$ is small. But it has
an angular dependence different from the previous contribution.
Figure \ref{onetwocrossings}.a shows that there is a phase factor
either for the incoming or the outgoing beam, so that instead of
(\ref{C1}), we have for this second contribution, usually called
$C_2$,

\be
 \overline{ \delta  T_{ab} \delta T_{a'b'}}^{C_2}= {2 \over 3 g}
\overline{
T}\hspace{-.1cm}_{ab}\hspace{.1cm}\hspace{-.2cm}^2\hspace{.1cm}
(\delta_{aa'} + \delta_{bb'}) \ \ . \label{C2}  \ee
The factor $2/3$ results from an integration over the position of
the quantum crossing \cite{AkkMon}. A sum over all incoming and
outgoing channels gives for this contribution~:

\begin{figure}[htb]
\centerline{ \epsfysize 7cm \epsffile{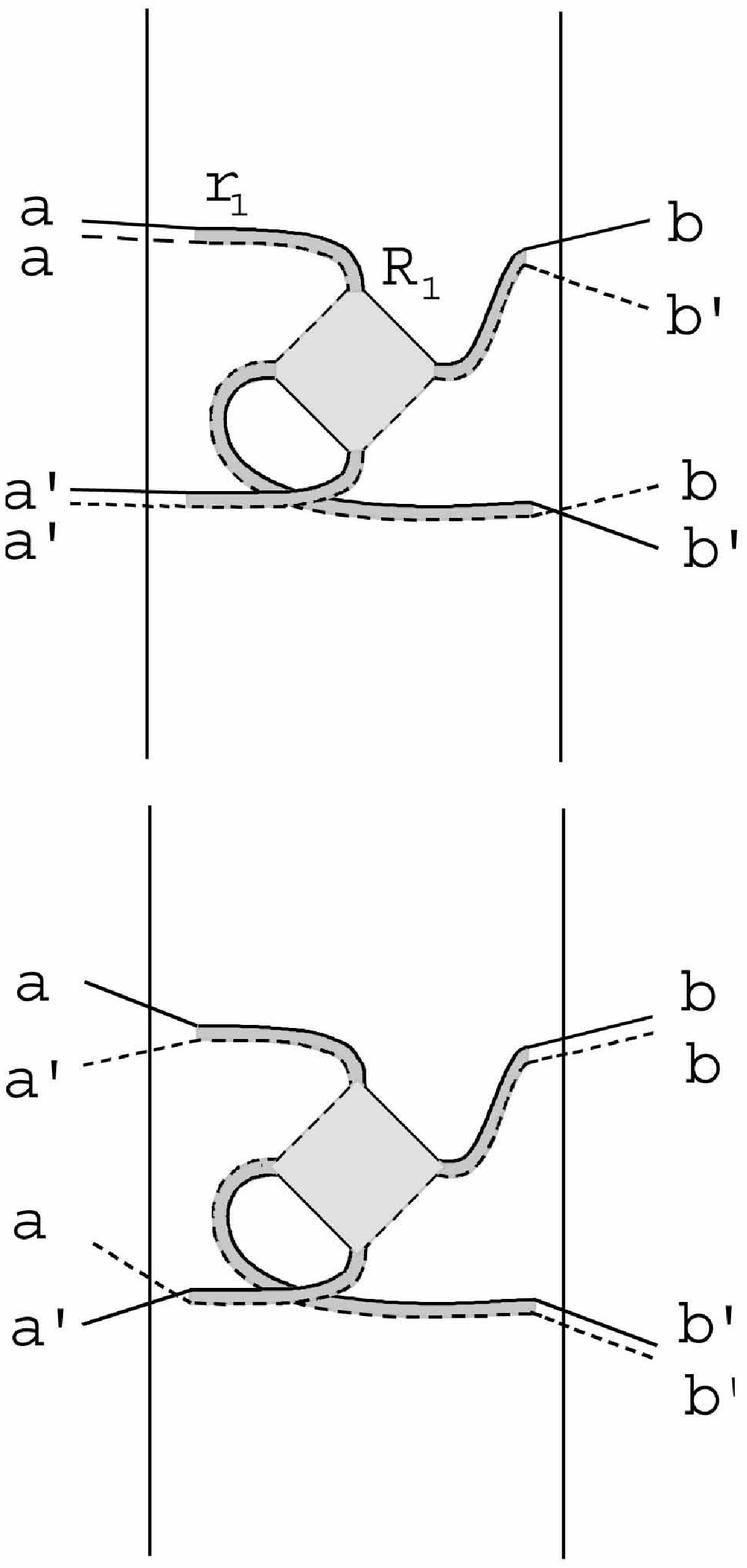}
\hspace{1cm} \epsfysize 7cm \epsffile{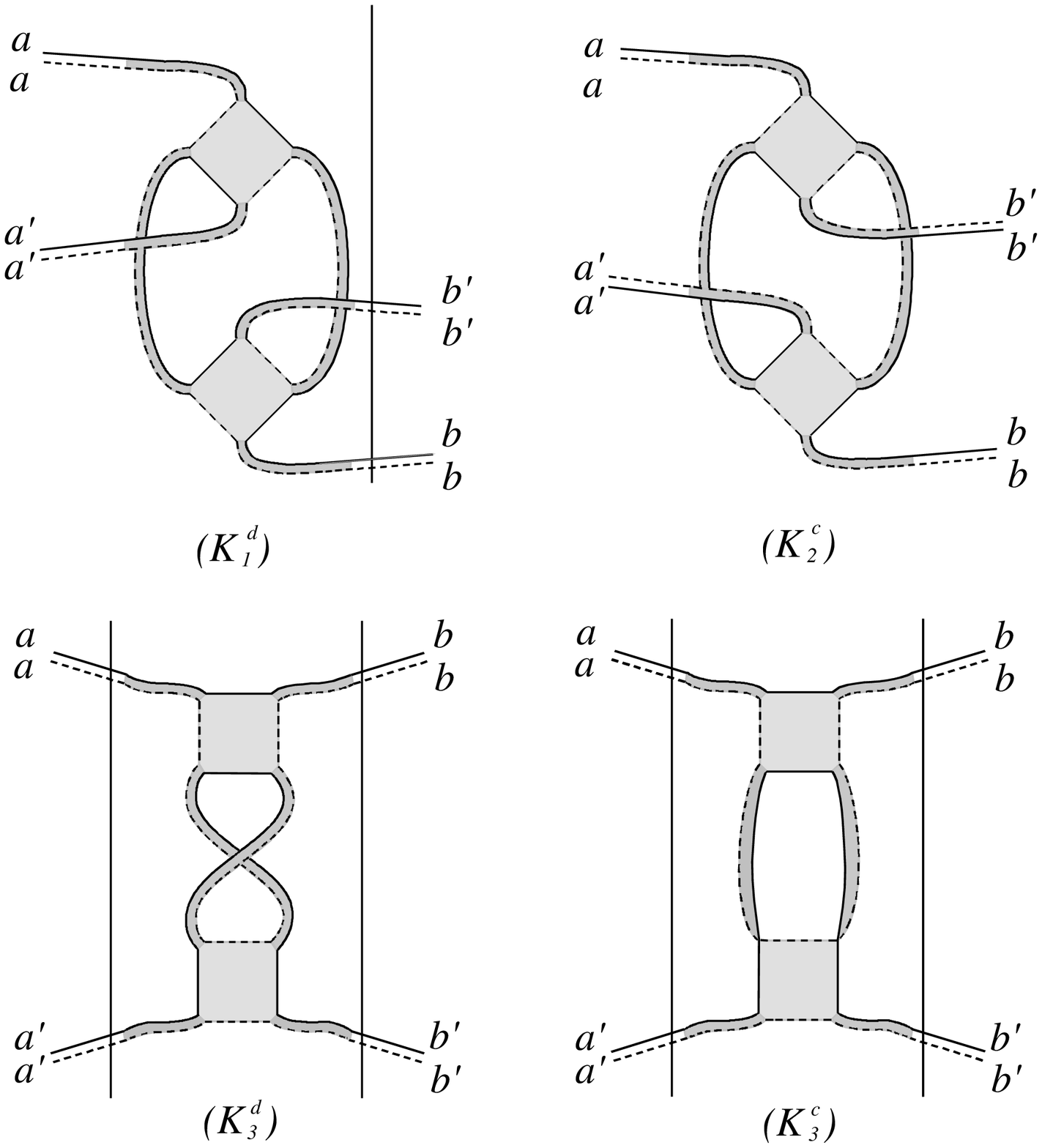}}
\hspace{1.5 cm} (a) \hspace{6 cm } (b) \caption{\em Schematic
representation of the $C_2$ and $C_3$ contributions to the
correlation function $\overline{ \delta T_{ab} \delta T_{a'b'}}$.
$\kappa_1^d$ and $\kappa_3^d$ are Diffuson contributions.
$\kappa_2^c$ and $\kappa_3^c$ are Cooperon contributions.}
\label{onetwocrossings}
\end{figure}

\be \overline{ {\delta g}^2}=\sum_{aa'bb'} {2 \over 3 g}
\overline{
T}\hspace{-.1cm}_{ab}\hspace{.1cm}\hspace{-.2cm}^2\hspace{.1cm}
(\delta_{aa'} + \delta_{bb'})=
 {4 g \over 3 M} \ll 1 \ \ . \ee
Still this contribution cannot explain the amplitude of the
observed conductance fluctuations, since it vanishes in the large
$M$ limit. Let us consider the next contribution shown on figure
\ref{onetwocrossings}.b, with two quantum crossings. We see that
this so-called ${C}_3$ contribution is  smaller than the first one
by a factor $1/g^2$~:

\be
 \overline{ \delta T_{ab} \delta T_{a'b'}}^{C_3}= {2 \over 15 g^2}
 \,
\overline{
T}\hspace{-.1cm}_{ab}\hspace{.1cm}\hspace{-.2cm}^2\hspace{.2cm}
\ee
The $2/15$ factor comes from the integration over the positions of
the two quantum crossings \cite{AkkMon} (see next section). This
term has {\it no angular dependence}, so that the summation over
channels contains now $M^4$ terms and the contribution to the
conductance fluctuations is

\be \overline{ {\delta g}^2}= {2 \over 15} \ \ . \ee The
fluctuations are universal, independent on the strength of
disorder. In summary, they are universal since the corresponding
correlation function is constructed with two conductances and two
quantum crossings, giving $g^2 \times 1/g^2 \simeq 1$. The
contributions with  no crossing or one crossing cancel because of
angular dependences. The next terms with $n$ crossings  are
negligible, of order $1/g^{n-2}$. In optics when one considers a
speckle pattern, the first contribution is the most important, and
the one crossing and two crossings contributions are very
difficult to observe (They can be observed since, although very
small, they are a different angular dependence,  and also
different temporal or frequency dependences \cite{AkkMon}). In
electronics, only the third contribution with two crossings  is
important after summation over incoming and outgoing channels. In
summary, $\sum_{ab} T_{ab}$ has much smaller fluctuations than
$T_{ab}$. We have now a simple recipe to evaluate average
quantities or correlations functions~: each quantum crossing gives
a factor $1/g$.

\subsection{Amplitude of the conductance fluctuations}

In order to calculate quantitatively the conductance fluctuations,
their dependence on geometry or external parameters, we must
analyze more precisely the structure of the paired trajectories in
figure \ref{onetwocrossings}.b. In addition to the crossings,
there is a loop. And we have to integrate on the distribution of
loops, like for the weak localisation correction. In contrast with
the weak localization correction, this loop is formed by two
crossings instead of one. So here for a given position of one
crossing, we have to integrate on the position of the second
crossing. Since this second crossing is necessarily along the loop
formed by the two crossings, the integration over the position of
the second crossing gives a volume element proportional to the
length of the loop. For a trajectory of length $v_F t$, the volume
is $v_F t {\lambda_F^{d-1}} $. Moreover, $P(t)$ contains
two-Diffusons and two-Cooperons contributions. A careful
examination of the possible crossing and trajectories shows the
only possible diagrams shown on figure \ref{onetwocrossings}.
Taking into account their degeneracy \cite{AkkMon}, we obtain an
expression which is as simple as the weak localization
correction~:

 \begin{equation}
\displaystyle  \overline{\delta G^2}   =  6 {s^2  } \left({e^2
\over h} \right)^2  \int_0^\infty t \ [P_{cl}(t)+P_{int}(t) ]
e^{-{ t/\tau_\phi}}{d
t \over \tau_D^2}\\
 \label{UCF} \end{equation}
There is an equal contribution of loops with Diffusons or
Cooperons. In a magnetic field, the Cooperon contribution is
suppressed so that the variance is reduced by a factor $2$, as
seen on fig. \ref{mailly}.c.  This happens for the same magnetic
field $B_\phi$ as the destruction of the weak localisation
correction, that is for a  flux quantum through the system or
through $L_\phi^2$.

Incoherent processes not only destroy the Cooperon contribution
but also the Diffuson contribution. This could appear surprising
since we have seen that this Diffuson contribution corresponds to
classical diffusion, and therefore should not be phase sensitive.
However, this Diffuson contribution is not really the classical
contribution, since it is constructed by pairing trajectories
corresponding to {\it different} realizations of the system. If
there is a phase breaking event, it affects equally one amplitude
and its complex conjugate. But here, the phase breaking event may
affect differently the two amplitudes since they correspond to
different systems.
\medskip

Like we have done above for the weak localization correction
(\ref{WL}), we can now evaluate quite easily     $\overline{\delta
G^2} $ given by (\ref{UCF})        for different geometries from
the corresponding expression of the return probability $P(t)$. Let
us do it for a quasi-$1d$ mesoscopic wire. $P(t)$ is given by

$$P(\q,t)=\sum e^{-D q^2 t} $$
where $q= n \pi /L$ is quantized by the Dirichlet boundary
conditions corresponding to a perfectly connected wire (no zero
mode, see section \ref{FSBC}). Inserting  $ P(t)$ in (\ref{UCF}),
we get \be \overline{\delta g^2}= 6  \sum_{n>0} {1 \over \pi^4
n^4}= {2 \over 15 }  \ \ . \ee
 Let us remark that the choice of the
boundary conditions is very important. If the system were closed
or poorly connected, corresponding to Neumann boundary conditions,
then the contribution of the zero mode would lead to a divergence,
or at least non universality of the fluctuations.

In the macroscopic limit, when $L_\phi \gg L$, we can treat the
system as infinite, replace $P(t)$ by its dependence (\ref{Pdetd})
for an infinite system $(\tau_D /4 \pi t)^{d/2}$ and multiply by
the exponential decay $e^{- t /\tau_\phi}$ in the integral. Then
we obtain~:

\be \overline{\delta g^2} \propto \left( {L_\phi \over L}
 \right)^{{4-d \over 2}} \ \ . \label{ucfred} \ee
 We are not surprised by this
result. It is exactly the one anticipated from our simple argument
(\ref{gLcL}) treating the fluctuations as due to incoherent
contributions of correlated regions of size $L_\phi$.

Finally, it is quite easy to notice, that  correlations functions
of $n$ conductances imply $2 n - 2$ crossings. Therefore the
$n^{th}$ cumulant of the conductance distribution is of order
$g^n/g^{2n-2} \propto 1/g^{n-2}$. It vanishes for $n > 2$ in the
metallic limit $g \rightarrow \infty$, so that  the conductance
distribution is indeed gaussian.

\section{Diffusion on graphs and spectral determinant}

The calculation of the weak localization correction or of the
conductance fluctuations, as well as of other thermodynamical
quantities like orbital magnetic susceptibility
\cite{AkkMon,graphs} can be extended   to the case of any
structure -- called a network -- made of quasi- one-dimensional
diffusive wires. First, we note that the quantities of interest,
like the weak localization correction (\ref{WL}) or the
conductance fluctuations (\ref{UCF}), have  the same structure~:

\begin{equation} \int t^\alpha P(t) e^{-\gamma t} dt \ \ ,
\end{equation}
 where   $\gamma= 1/\tau_\phi$. From (\ref{partition}), the time integral of $P(t)$ can be straightforwardly written in terms of a quantity
called the {\it spectral determinant} $S(\gamma)$~:
\begin{equation}
  \int_0^\infty dt  P(t) e^{- \gamma t}
 = \sum_n \frac{1}{E_n + \gamma}
= \frac{\partial}{\partial \gamma} \ln { S} (\gamma) \ \ ,
\label{diff.spectral}
\end{equation}
where $S(\gamma)$ is, within a multiplicative constant independent
of $\gamma$~:
\begin{equation}
{S} (\gamma) = \prod_n (\gamma + E_n) \ \ ,  \label{SD}
\end{equation}
 $E_n$ being the eigenvalues of the diffusion equation (\ref{schrodinger}).
Using standard properties of Laplace transforms, the above time
integrals can be rewritten in terms of the spectral determinant,
so that the  weak localization to the conductivity and the
conductivity fluctuations respectively read~:
\begin{eqnarray}
\Delta \sigma &=&                  -    \ 2 s {e^2 \over  h } {D
\over \Omega}\ \ \ \   {\partial \over \partial \gamma}\ln
S_{int}(\gamma)  \label{WL2}\\
\langle \delta \sigma^2 \rangle &=& - \ 6 s^2   {e^4 \over
h^2}{D^2 \over \Omega^2} \  {\partial^2 \over \partial
\gamma^2}\left[\ln S_{cl}(\gamma) + \ln S_{int}(\gamma)\right] \ \
, \label{UCF2}
\end{eqnarray}
where $S_{cl}$ and $S_{int}$ are the spectral determinants
associated respectively to the diffusion equation for the Diffuson
and the Cooperon. $\Omega$ is the total volume of the system.
These expressions are quite general, strictly equivalent to
expressions (\ref{WL},\ref{UCF}). Their interest is that, on a
network, the spectral determinant takes a very simple form.
 By solving the diffusion equation (\ref{diff2}) on
each link, and then imposing Kirchoff type conditions on the nodes
of the graph with $N$ nodes,  the problem can be  reduced to the
solution of a system of $N$ linear equations relating the
eigenvalues at the $N$ nodes. Let us introduce the $N \times N$
matrix $M$~:
\begin{equation}
M_{\alpha \alpha} = \sum_\beta \coth(\eta_{\alpha \beta}) \ \ , \
\ M_{\alpha \beta} = - \frac{e^{i \theta_{\alpha \beta}}} {\sinh
\eta_{\alpha \beta}} \label{matrixM}
\end{equation}
The sum $\sum_\beta$ extends to all the nodes $\beta$ connected to
the node $\alpha$; $l_{\alpha \beta}$ is the length of the link
between $\alpha$ and $\beta$. $\eta_{\alpha \beta} = l_{\alpha
\beta}/L_\phi$. The off-diagonal coefficient $M_{\alpha \beta}$ is
non zero only if there is a link connecting the nodes $\alpha$ and
$\beta$. $\theta_{\alpha \beta} = (4 \pi /\phi_0)\int_\alpha
^\beta A.dl$ is the circulation of the vector potential between
$\alpha$ and $\beta$. $N_B$ is the number of links in the graph.
It can then be shown that the spectral determinant takes the very
convenient form~\cite{AkkMon,graphs}~:

\begin{equation}
{S} = \left ( \frac{L_\phi}{L_0} \right ) ^{N_B -N} \ \prod
_{(\alpha \beta)} \sinh \eta_{\alpha \beta} \ \det M
\label{det.spectral.Q1D} \end{equation}
 $L_0$ is an arbitrary length independent of $\gamma$ (or $L_\phi$).
We have thus transformed the spectral determinant which is an
infinite product in a finite product related to $\det M$. Using
(\ref{det.spectral.Q1D}), mesoscopic quantities
(\ref{WL2},\ref{UCF2}) can be easily predicted for any geometry of
diffusive networks \cite{AkkMon,graphs}.

\medskip

\section{Interaction effects}

Until now we have not considered the role of electron-electron
interactions. They turn out to give small corrections to transport
quantities like the average conductivity, but they play an
important role to understand thermodynamic properties like
persistent currents \cite{AkkMon}. Moreover, until now we have
introduced by hand a phase coherence time $\tau_\phi$ (or length
$L_\phi$). We wish now to understand the microscopic origin for
the loss of quantum coherence. This phase coherence is limited by
the interactions with other degrees of freedom, in particular
other electrons through their mutual interaction. We want to
discuss now how e-e interactions break phase coherence.
\medskip

On one hand, interaction effects can be considered as negligible.
We know from Landau theory of Fermi liquids that in an interacting
electron gas, free particles have simply to be replaced by {\it
quasiparticles} which are dressed objects, screened by the could
of other electrons.  These quasiparticles have a long lifetime
which diverges when approaching the Fermi level. From Landau, we
know that
\be {1 \over \tau_{ee}(\ep) }\propto \ep^2
\label{landau}  \ee
where $\ep=E-E_F$ is the energy of the quasiparticle measured from
the Fermi level. On the other hand, in a disordered metal,
electrons move diffusively, that is very slowly. They spend long
time close to each other. Qualitatively, we can expect that
diffusion somehow enhance the effect of interactions. We may ask
if and how expression (\ref{landau}) is changed because of the
diffusion. Moreover, since the diffusive motion is dimensionality
dependent, the modified lifetime should also depend on this
dimensionality.

\medskip

Interaction between electrons is expected to have two major
effets~:
\begin{itemize}

\item First, each electron is not only sensitive to the disordered
potential but also to the fluctuations of the electronic density
due to other electrons. This additional fluctuating potential
modifies the position of the energy levels, especially near the
Fermi level. So we may expect a modification in the repartition of
the energy levels, that is a change of the density of states near
the Fermi level. We shall show that the density of states exhibits
a decrease, the so-called "Altshuler-Aronov" anomaly. This
reduction of the density of states is accompanied by a reduction
of the conductivity.
\medskip

\item Since the e-e interaction is a inelastic process, each
quasiparticle has a finite lifetime which  limits the phase
coherent properties like weak localization, since the coherence
between time-reversed trajectories in necessarily limited by this
lifetime. \end{itemize}

\subsection{Screening}

In order to describe interaction effects, let us start with a few
reminders about the screening of interaction. The bare Coulomb
interaction potential is $U_0(R)=e^2/R$, that is in $3d$,
$U_0(q)=4 \pi e^2 /q^2$. The screened Coulomb interaction $U(q)$
is given by \be U(q)={U_0(q) \over 1 + \chi_0(q) U_0(q)} \ee
where, in the small $q$ limit (we are interested in the diffusive
regime where $q l_e \ll 1$), the susceptibility $\chi_0(q)$ is the
Pauli susceptibility, that is the density of states $2 \rho_0$.
Therefore, the screened interaction is given by
\be U(q)= {4 \pi
e^2 \over q^2 + \kappa^2} \qquad , \qquad U(R)={ e^2  \over R}
e^{- \kappa R}  \ \ ,  \ee
where the Thomas-Fermi vector $\kappa$ (inverse screening length)
is
\be \kappa^2 = 8 \pi e^2 \rho_0 \ \ .  \ee
In the diffusive limit $q l_e \ll 1$, the screened interaction can
by approximated by
\be U(q)={4 \pi e^2 \over \kappa^2}= {1 \over \chi_0}= {1 \over 2
\rho_0} \qquad , \ \mbox{that is} \qquad U(R)={1 \over 2 \rho_0}
\delta(R) \label{Ustat}  \ \ . \ee
On the scale of diffusion, the
screened interaction can be considered as a local interaction.

However, it turns out that screening is not instantaneous since
electrons have to diffuse to screen a local charge. Therefore the
interaction is actually frequency dependent. This dynamical
screening is described by the frequency dependent susceptibility
$\chi_0(q,\omega)$ which accounts for the dynamical charge
reorganization~:
\be \chi_0(q,\omega)=2 \rho_0 \, {D q^2 \over - i \omega + D q^2}
\ \ . \ee
Therefore eq. (\ref{Ustat}) becomes
\be U(q,\omega)= {1 \over \chi_0}= {1 \over 2 \rho_0} {- i \omega
+ D q^2 \over D q^2} \ \ .  \label{Uchi0} \ee

\subsection{Density of states anomaly}

 As we have done above, we shall
avoid technicalities of diagrammatic theory, and try to get the
important results from qualitative arguments. Although we have not
elaborated on the theory of Green's functions, let me remind you
at least that the Green's function is related to the density of
states by
\be \rho(\ep)={-1 \over \pi \Omega }\int  \mbox{Im}
G_\ep(\rr_0,\rr_0) d \rr_0 \ \  . \ee
As we have seen in section \ref{CPDE}, $G(\rr_0,\rr_0)$ is the sum
of contributions from all closed trajectories from $\rr_0$ to
$\rr_0$ (figure \ref{dosAA}.a). All these amplitudes have
different different and random phases and their contribution
cancels in average. What remains is the contribution of short
trajectories, giving an average density of states $\overline{
\rho(\ep)}=2 \rho_0$.

How can we effects of diffusion + interactions appear on the
density of states? The non-interacting density of states (or
Green's function) is a single electron property, and therefore
involve single trajectories. In the presence of electron-electron
interaction, each electron trajectory can be paired with the
trajectory of a second electron, with which it interacts. Then by
pairing these two trajectories, we can construct a Diffuson. More
precisely, we pair an amplitude corresponding to one electron to
the conjugate amplitude corresponding to another electron. Their
interaction  is represented by a wiggly  line in Figure
\ref{dosAA}.b. There are actually two possible contributions,
depending on the position of the interaction line. They are
nothing but the Hartree and the exchange (Fock) contributions. We
can conveniently separate the "diagrams" in three different
parts~:
\begin{itemize}
\item  A short range part close to the point $\rr_0$. It
ressembles somehow to a quantum crossing, with a dephasing between
the three trajectories,

\item two long ranged Diffusons,

\item an interaction region between $\rr$ and $\rr'$.
\end{itemize}

\noindent So, we can easily construct the structure of this
additional contribution to the density of states~:

\begin{figure}[ht]
\centerline{ \epsfxsize 10cm \epsffile{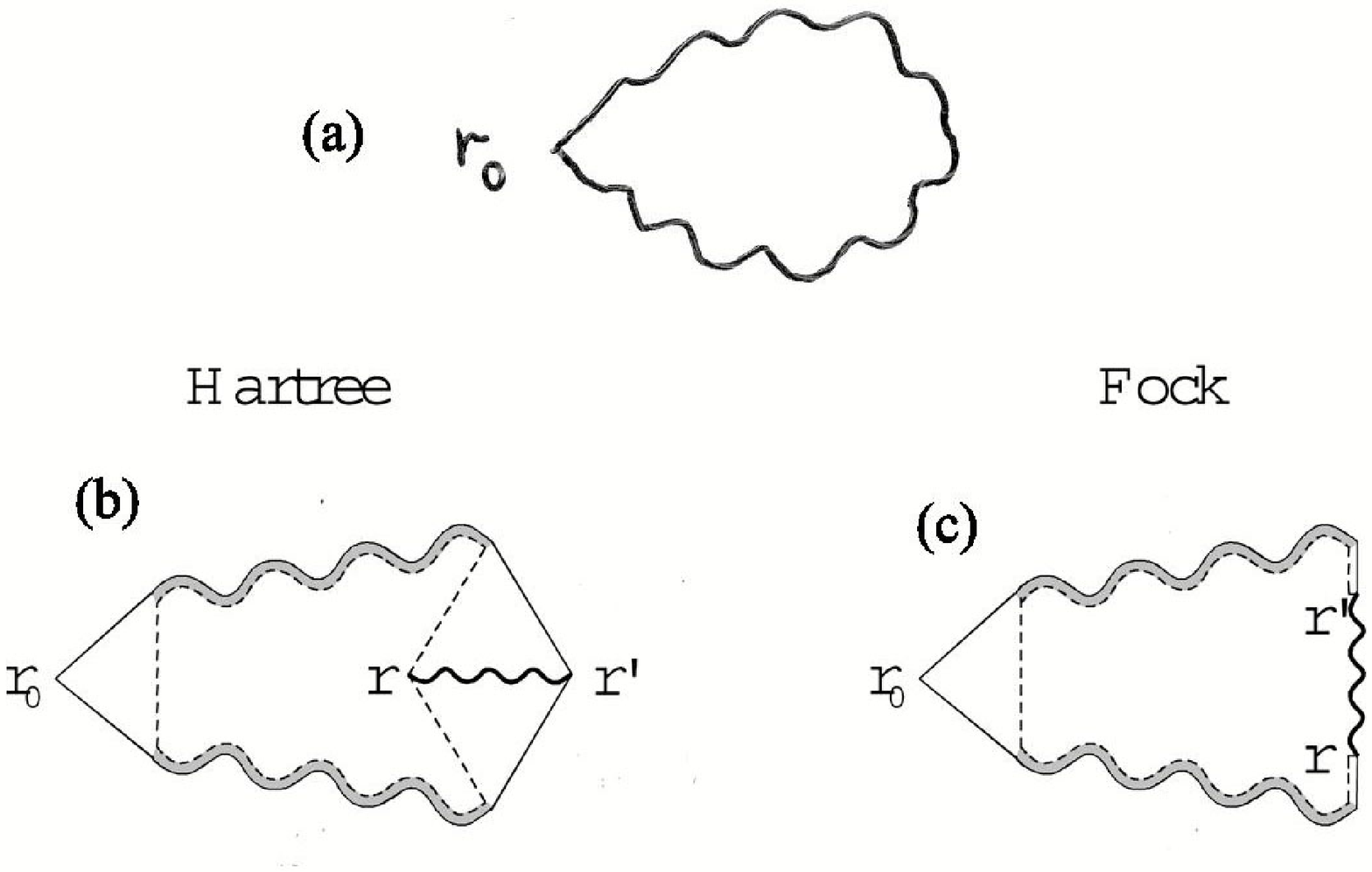} } \centerline{
\epsfxsize 10cm \epsffile{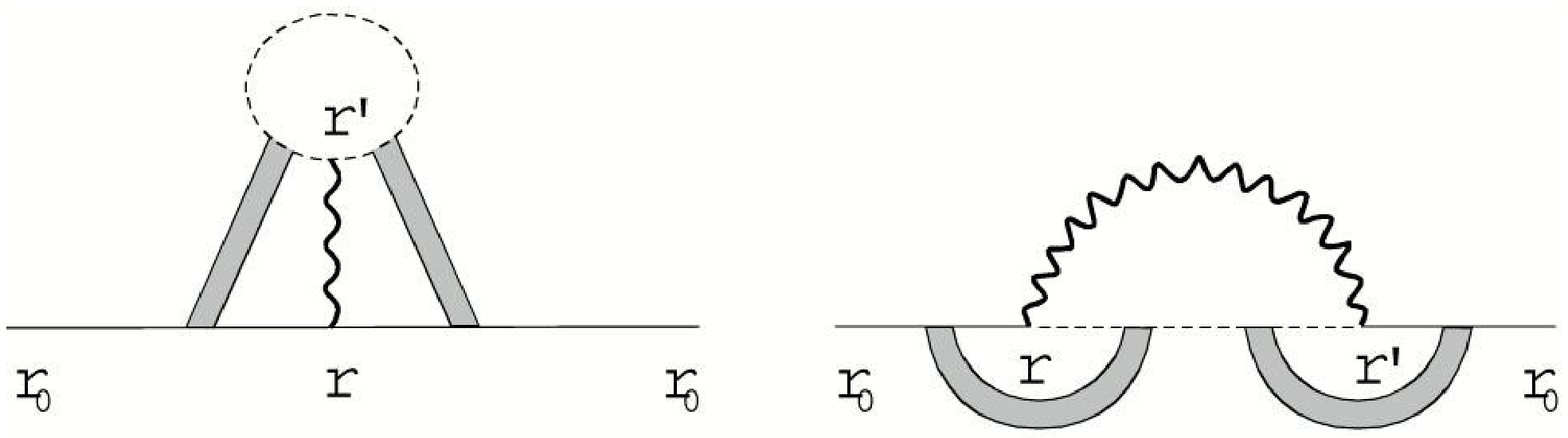} } \caption{\em a) Diagram
for the non-interacting density of states. After disorder
averaging, the contribution of diffusive trajectories vanishes
because of their random phase. Hartree  (b) and exchange (c Fock)
diagrams for the  density of states anomaly. These contributions
survive disorder averaging since they contain paired trajectories.
Upper~: schematic representation exhibiting clearly the three
regions, a short range region, diffusive trajectories and the
interaction region. Bottom~: usual diagrammatic representation.
These two representations are equivalent. } \label{dosAA}
\end{figure}

\be {\delta \rho \over \rho_0} \simeq - {\lambda_\rho \over g}
\int^{\hbar /\ep} P(t) \ { dt \over \tau_D}  \ \ .
\label{dosanomaly} \ee
$\lambda_\rho$ is a dimensionless parameter which describes the
strength of the interaction. It contains both the Hartree and
exchange contributions. The $1/g$ reduction comes from the quantum
crossing. And the distribution of loops formed by the paired
trajectories is given by the return probability $P(t)$. For an
energy $\ep$, the two amplitudes must be taken at different
energies $\omega$ and $\omega-\ep$. Therefore the two trajectories
can stay in phase only during a time $\Delta \simeq \hbar /\ep$,
so that the upper cutoff in the integral is $\hbar /\ep$.

 Here our
aim is simply to present the structure of the result without
entering into details. Another qualitative derivation is given in
\cite{Aussois}, a detailed discussion is proposed in
\cite{AkkMon}, and the original calculation is done  in
\cite{AltshulerAronov}. We see that the amplitude and the
structure of this correction to the density of states looks very
similar to that of the weak localization correction (\ref{WL}),
except that the upper cutoff is not $\tau_\phi$ but $\hbar /\ep$.
Unlike for the weak localization, this correction depends on the
classical return probability, so that it is not suppressed by a
magnetic field \cite{Cooperon}.

The form (\ref{dosanomaly}) is approximate. A more sophisticated
calculation replaces the upper cutoff by a Fourier transform.
Moreover since $g \propto \rho_0 \Omega /\tau_D$, eq.
(\ref{dosanomaly}) becomes

\be {\delta \rho(\ep) } = -{\lambda_\rho \over \pi \Omega}
\int_0^\infty P(t) \cos \ep t \, dt   \ \ . \label{dosanomaly2}
\ee
From the expression (\ref{Pdetd}) of $P(t)$, we obtain the energy
dependence of the density of states anomaly~:
 \be
\displaystyle \delta \rho (\ep)  \propto -{\lambda_\rho \over D
\Omega} \left\{
\begin{tabular}{lll}
$\displaystyle{ L_{\ep} }$ &  & $quasi-1d$ \\
\\
$\displaystyle{  \ln {L_\ep \over l_e} }$ &  & $d=2$ \\
\\
$\displaystyle{ {1 \over l_{e}} - {1 \over L_\ep}}$ &  & $d=3$%
\end{tabular}
\right. \label{arevoir1} \ee
where the characteristic length $L_\ep$ is $L_\ep=\sqrt{\hbar D
/\ep}$. The structure of this correction related to $P(t)$ is
similar to the weak localization correction (\ref{sigmad}), except
that the characteristic length $L_\phi$ has been replaced by
$L_\ep$. This reduction of the density of states can be observed
experimentally by tunnel measurements, since it reflects as an
anomaly in the voltage dependence of the tunnel conductance $G_t$.
At zero temperature, the relative correction to the tunnel
conductance is given by
\be {\delta G_t \over G_t} ={\delta \rho(\ep= eV) \over \rho_0} \
\ . \ee

At finite temperature, it is not difficult to generalize the
expression (\ref{dosanomaly2}) of the density of states anomaly as
\be \delta \rho(\ep, T)= - \int f'(\ep-\omega) \delta \rho(\omega)
d \omega \ \ , \ee where $f'$ is the derivative of the Fermi
function. After a Fourier transform, we find

\be {\delta \rho(\ep,T) } = -{\lambda_\rho \over 2 \pi \Omega}
\int_0^\infty R_T(t) P(t) \cos \ep t \, dt   \ \ ,
\label{dosanomaly3} \ee
where the thermal function $R_T(t)$ is given by $R_T(t)= \pi T  t
/\sinh \pi T t$. The temperature dependence of the tunnel
conductance anomaly, also called zero-bias anomaly is
 \be {\delta G_t(V,T) \over G_t}= -{1 \over \rho_0} \int \delta
\rho(\ep,T) f'(\ep- eV) d \ep \ \ , \ee or, after a Fourier
transform~:

\be {\delta G_t(V,T) \over G_t}= -{\lambda_\rho \over 2 \pi \rho_0
\Omega} \int_0^\infty R^2_T(t) P(t) \cos e V t \, dt   \ \ ,
\label{dosanomaly4} \ee
This correction has been measured for various systems with
different dimensionalities and the $1/\sqrt{V}$, $\ln V$, and
$\sqrt{V}$ predicted by (\ref{arevoir1},\ref{dosanomaly4})
respectively in $1$, $2$ and $3$ dimensions have been observed
\cite{tdos}.

\subsection{Correction to the conductivity}

Taking into account the interaction between electrons  leads also
to a reduction of the conductivity. Without going into the details
of the calculations, we can argue that this reduction is a
consequence of the correction to the density of states. Both
effects result from the scattering of an electron by the charge
fluctuations induced by disorder. The temperature dependence of
the conductivity $\sigma(T)$  is related to its energy dependence
at $T=0K$ by $\sigma(T)=-\int f'(\ep) \sigma(\ep) d\ep$, where
$f'(\ep)$ is the derivative of the Fermi function. Since the
conductivity is proportional to the density of states (Einstein
relation), we expect that the density of states anomaly leads to a
correction of the conductivity given by
\be {\delta \sigma(T) \over \sigma_0} =    \int d\ep
\left(-{\partial f \over
\partial \ep}\right){\delta \rho(\ep,T) \over \rho_0} \ \ ,
\ee
where $\sigma_{0}$ is the  Drude conductivity
(\ref{sigmaeinstein}). For a static interaction, the density of
states correction $\delta \rho(\ep,T)$ is given by
(\ref{dosanomaly3}). Upon Fourier transforming, we have

\begin{equation}
\displaystyle {\delta \sigma (T) } = -\lambda_\sigma \ \left(e^2 D
\over \pi \Omega\right)  \int_0^\infty  R_T^2(t) P(t)
 dt
\label{delsigint} \ee
where $\lambda_\sigma$ is a parameter which depends on the
interaction \cite{lambdas}. Using the expression (\ref{Pdetd})  of
$P(t)$, we obtain the temperature dependence of the correction to
the conductance

 \be
\displaystyle \delta g (T)  \propto -{\lambda_\sigma } \left\{
\begin{tabular}{lll}
$\displaystyle{{ L_{T} \over L} }$ &  & $quasi-1d$ \\
\\
$\displaystyle{  \ln {L_T \over l_e} }$ &  & $d=2$ \\
\\
$\displaystyle{ {L \over l_{e}} - {L \over L_T}}$ &  & $d=3$%
\end{tabular}
\right. \label{arevoir2} \ee
where the thermal length $L_T$ is defined by $L_T=\sqrt{\hbar D /
T}$. In $2d$, the temperature dependence is logarithmic like the
weak localization correction. Unlike the weak localization
correction, this correction to the conductivity is not sensitive
to a magnetic field. Therefore both corrections can be separated
experimentally  by the application of a magnetic field.

\subsection{Lifetime of quasiparticle}

 Consider a Fermi sea and inject a quasiparticle in a state $|\alpha
 \rangle$ with energy $\ep_\alpha$
 above the Fermi sea. It interacts with another particle
($|\gamma \rangle, \ep_\gamma$) and the final state consists in
two quasiparticles ($|\beta \rangle, \ep_\beta$) and ($|\delta
\rangle, \ep_\delta$) above the Fermi sea (figure \ref{tauee}).
Energy conservation implies $\ep_\alpha + \ep_\gamma=
\ep_\beta+\ep_\delta$. The lifetime of the state $|\alpha \rangle$
is given by the Fermi golden rule and it is related to the matrix
element of the interaction~:

\begin{figure}[ht]
\centerline{ \epsfxsize 6cm \epsffile{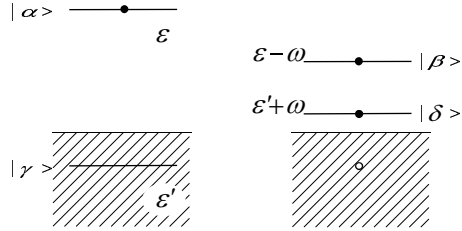} } \caption{\em A
quasiparticle in a state $|\alpha\rangle$ of energy
$\ep_\alpha=\ep$ interacts with another quasiparticle  $|\gamma
\rangle$ of energy $\ep_\gamma=\ep'$ in the Fermi sea. The final
state is made of two quasiparticles above the Fermi sea and one
hole.} \label{tauee}
\end{figure}

\be {1 \over \tau_\alpha}= 2 \pi s \sum_{\beta\gamma\delta}
|\langle \alpha \gamma | U | \beta \delta \rangle|^2
\delta(\ep_\alpha + \ep_\gamma- \ep_\beta-\ep_\delta) \ \ ,
\label{FGR1} \ee
with the constraint that $\ep_\gamma <0$, $\ep_\beta >0$ and
$\ep_\delta >0$. If the matrix element does not depend on
energies, we see immediately that the Landau $\ep^2$ dependence
comes simply from energy constraints~:  basically three final
states can be chosen in a range of energy $\ep$, with the
constraint of energy conservation, whence the $\ep^2$ dependence.
Indeed the matrix element of the interaction is energy independent
in the ballistic case, but this is to true anymore in the
diffusive case. The goal of the following pages is to calculate
the energy dependence of the litetime in the diffusive regime.
Since we do not specify a given state $|\alpha \rangle$, eq.
(\ref{FGR1}) can be rewritten for the lifetime at a given energy
$\ep$

\be {1 \over \tau_{ee}(\ep)}= {1 \over \nu_0 } \sum_\alpha {1
\over \tau_\alpha} \delta(\ep -\ep_\alpha)= { 2 \pi s \over \nu_0}
\sum_{\alpha\beta\gamma\delta} |\langle \alpha \gamma | U | \beta
\delta \rangle|^2 \delta(\ep_\alpha + \ep_\gamma-
\ep_\beta-\ep_\delta)  \delta(\ep -\ep_\alpha) \ \ , \label{FGR2}
\ee
where $\nu_0=\rho_0 \Omega$ is the density of states. By
introducing two energy integrals, it can be rewritten in the form

\be {1 \over \tau(\ep)}= { 2 \pi s \over \nu_0 } \int_0^\ep d
\omega \int_{- \omega}^0 d\ep' \sum_{\alpha\beta\gamma\delta}
|\langle \alpha \gamma | U | \beta \delta \rangle|^2 \delta(\ep -
\ep_\alpha) \delta(\ep'- \ep_\gamma) \delta(\ep-\omega-
\ep_\beta)\delta(\ep'+\omega-\ep_\delta) \ \ , \label{FGR3} \ee
where $\ep>0$, $\ep'<0$ and $\ep-\omega>0$,  $\ep'+\omega>0$ are
respectively the two  energies of the initial states  and of the
final states. If the matrix element is energy independent, we
trivially recover the $\ep^2$ dependence. Upon averaging over
disorder, the lifetime has the form

\be
\displaystyle {1 \over \tau_{ee}(\ep )}= {4 \pi \nu_0^3}
\int_0^\ep \omega W^2(\omega) d \omega \label{taugenW2} \ee
 with
 \be W^2(\omega)={1 \over \nu_0^4}
\overline{\sum_{\alpha\beta\gamma\delta} |\langle \alpha \gamma
|U|\beta \delta\rangle|^2 \delta(\ep-\ep_\alpha)
\delta(\ep'-\ep_\gamma) \delta(\ep-\omega-\ep_\beta)
\delta(\ep'+\omega- \ep_\delta)} \label{W2} \ee
The characteristic matrix element  $W(\omega)$  depends
 only on energy transfer $\omega$, but  neither on $\ep$, nor on $\ep$.

The matrix element  $\langle \alpha \gamma | U |\beta \delta
\rangle$ can be evaluated on the basis of eigenfunction of the non
interacting particles and reads
 \be \langle \alpha \gamma | U |\beta \delta \rangle=
\int d\rr_1 d\rr_2 \phi_\alpha^*(\rr_1) \phi_\gamma^*(\rr_2)
\phi_\beta(\rr_1) \phi_\delta(\rr_2)U_\omega(\rr_1-\rr_2) \ \ ,
\ee
where $U_\omega(\rr)$ is the dynamically screened potential. The
combination of wave functions and $\delta$ function can be
rewritten in terms of Green's functions
\be -{1 \over \pi }\mbox{Im} G(\rr,\rr')= \sum_\alpha
\phi_\alpha^*(\rr) \phi_\alpha(\rr') \delta(\ep - \ep_\alpha) \ \
, \ee
so that
  $W^2(\omega)$  can be rewritten in the form

\begin{eqnarray}
W^2(\omega)&=& {1 \over   \nu_0^4 \pi^4 }\int d\rr_1 d\rr_2
d\rr'_1 d\rr'_2 U_\omega(\rr_1- \rr_2)U_\omega(\rr'_1-\rr'_2)
\nonumber \\
&\times&\overline{\mbox{Im}G_\ep(\rr_1,\rr'_1)
\mbox{Im}G_{\ep-\omega}(\rr'_1,\rr_1)}\ \
\overline{\mbox{Im}G_{\ep'}(\rr_2,\rr'_2)
\mbox{Im}G_{\ep'+\omega}(\rr'_2,\rr_2)}
\label{W2GGa}\end{eqnarray}
where the average the the product of four Green's functions has
been decoupled into the product of two average values. By pairing
Green's functions, it is possible to show that the average product
contains a long-range part related to the probability $P$ (see eq.
\ref{proba} and \cite{AkkMon})

\be \overline{\mbox{Im}G_\ep(\rr,\rr')
\mbox{Im}G_{\ep-\omega}(\rr',\rr)}={\pi \rho_0}\,  \mbox{Re}
P(\rr,\rr',\omega) \ \ .  \label{ImGGrGa2'} \ee
We deduce
\begin{eqnarray}
W^2(\omega)&=&{1 \over \pi^2 \nu_0^2 \Omega^2 }\int d\rr_1 d\rr_2
d\rr'_1 d\rr'_2 U_\omega(\rr_1-\rr_2)U_\omega(\rr'_1- \rr'_2)
\nonumber \\
&\times&\mbox{Re}P_d(\rr_1,\rr'_1,\omega)\mbox{Re}P_d(\rr_2,\rr'_2,-\omega)
\label{W2q21}
\end{eqnarray}
or, upon  Fourier transformation~:
 \be W^2(\omega)={1 \over \pi^2 \nu_0^2 \Omega^2
}\sum_{\q \neq 0} |U(\q,\omega)|^2 [ \mbox{Re}P_d(\q,\omega)]^2  \
\ .  \label{W2q20}\ee
At the diffusion approximation, the dynamically screened potential
is given by (\ref{Uchi0}) so that
\be W^2(\omega)={1 \over 4 \pi^2 \nu_0^4 }\sum_{\q \neq 0} {1
\over \omega^2 + D^2 q^4} \ \ , \label{W2q2}
\end{equation}
which can be expressed in terms of the return probability  $P(t)$
\be W^2(\omega)={1 \over 4 \pi^2 \nu_0^4 } {1 \over \omega}
\int_0^\infty P(t) \sin \omega t\ dt \ \ . \label{W2q32} \ee
Finally,   the electronic lifetime (\ref{taugenW2}) is given by
\be \displaystyle {1 \over \tau_{ee}(\ep)}= {2 \over \pi \nu_0}
\int_0^\infty {P(t) \over t} \sin^2 {\ep t \over 2} dt
\label{taueezdet} \ee \medskip
For  a metal of volume $\Omega$, we can identify two different
regimes~:
\medskip

$\bullet$  $\ep \gg E_{c}$ where $E_c$ is the Thouless energy.
This corresponds to time scales  $t \ll \tau_{D}$. In this case an
electron described as a diffusive wave packet is insensitive to
the boundaries and behaves as in an infinite medium where,
according to (\ref{Pdetd}), $P(t)=\Omega/(4 \pi D t)^{d/2}$. We
obtain for the integral (\ref{W2q32})~: \cite{sumintegral}
\begin{equation}
W^2(\omega)={d c_d \over 16} {1 \over \nu_0^4 \omega^2 }\left(
{\omega \over E_c} \right)^{d/2} \ \ ,  \label{W2q33}
\end{equation}
so that  the electronic lifetime is equal to
\be
\begin{array}{c}
\\
\displaystyle {1 \over \tau_{ee}(\ep)}={\pi \over 2} c_d \Delta
\left( {\ep \over E_c} \right)^{d/2}
\\
\\
\end{array}
\ \ \  \ \ \ \ (\ep \gg E_c)
 \label{AAK} \ee
 where
  $c_1={\sqrt{2}/ \pi^2}$, $c_2={1 / 4
\pi^2}$, $c_3={\sqrt{2}/ 6 \pi^3}$. Such a behavior has been
indeed observed in silver wires  ($d=1$) for which $W^2(\omega)
\propto \omega^{-3/2}$ and $1/ \tau_{ee}(\ep)\propto \ep^{1/2}$,
although the measured prefactor came out to be larger than the
value predicted here \cite{Pierre99}.
\medskip

$\bullet$ The limit $\ep \ll E_c$, that is $t \gg \tau_{D}$,
corresponds to the ergodic regime in which the diffusive
electronic wave packet explores all the accessible volume
$\Omega$. Thus we would expect  $P(t)$ to be driven only by the
zero mode.  This is not so, because in expression (\ref{W2q2})
this mode has been removed in order to ensure electronic
neutrality. The excitation energy $\ep$ is smaller than
 $E_c$ and it is not possible to replace the sum (\ref{W2q2}) by an integral. In this
 limit, we obtain
\be W^2(\omega)= {a_d \over 4 \pi^6}   {\Delta^4  \over  E_c^2}
\propto {\Delta^2 \over g^2} \ \ , \ee
 where the coefficient $a_d$ is defined by the series
 \be a_d=\sum_{n_x,n_y,n_z} {1 \over (n_x^2+n_y^2+n_z^2)^2 } \ \ .\ee
The ratio  $E_c/\Delta$ is  the dimensionless conductance $g$.
 For  $\omega \ll E_c$,
the characteristic matrix element of the interaction is thus
energy independent and of  order  $\Delta /g$. The inverse
lifetime in this case is \cite{Sivan}
\be
\begin{array}{c}
\\
\displaystyle {1 \over \tau_{ee}(\ep )}= {a_d \over 2 \pi^5 }
\Delta \left( {\ep \over E_c} \right)^{2}
\\
\\
\end{array}
 \ \ \  \ \ \ \ (\ep \ll E_c) \ \ .
\label{Sivan} \ee

\subsection{Quasiparticle lifetime at finite temperature}

 In
Landau theory it is well-known that the quasiparticle lifetime at
zero energy $\ep=0$ and finite temperature $T$ is simply obtained
by replacing $\ep$ by $T$, so that it varies as $T^2$. The
diffusive case is more subtle. It turns out that in this case we
cannot simply substitute $\ep$ by $T$. This is wrong in low
dimension. Let us see why.

What is changed at finite temperature? The Fermi golden rule is
modified to account for Fermi factors. The condition of filled or
empty states has to be replaced by Fermi factors and eq.
(\ref{FGR3}) generalizes  as

\be {1 \over \tau_{ee}(\ep,T)}= 4 \pi \nu_0^3
\int_{-\infty}^\infty d \omega \int_{-\infty}^\infty d \ep'
F(\ep,\ep',\omega) W^2(\omega) \ee
where $F(\ep,\ep',\omega)$ is a combination of Fermi factors
$f_\ep=1/(e^{\beta \ep}+1)$~:
\be F(\ep,\ep',\omega)= f_{\ep'}
(1-f_{\ep-\omega})(1-f_{\ep'+\omega})
+(1-f_{\ep'})f_{\ep-\omega}f_{\ep'+\omega} \ . \label{Ft} \ee
 The first term in this expression is larger when $\ep >0$. It describes the decay
  of an electron-like  state above the Fermi level.
     The second term dominates when $\ep<0$ and describes the decay of a hole-like state
      into the Fermi sea.
For $\ep=0$, both terms are equal. Integrating upon $\ep'$, we
obtain
\be {1 \over \tau_{ee}(\ep,T)}= {4 \pi  \nu_0^3}
\int_{-\infty}^\infty d \omega \ \omega W^2(\omega) f_{\ep-\omega}
{e^{\beta \ep}+1 \over e^{\beta \omega } -1} \ \
.\label{taueeT1}\ee
This lifetime can also be obtained from the imaginary part of the
self-energy of a quasiparticle in the presence of a screened
interaction \cite{Abrahams81}. At zero  temperature, we recover
the result (\ref{AAK}).

\subsection{Quasiparticle lifetime at the Fermi level}
\label{sect.relaxE }

We now consider more specifically the lifetime of a quasiparticle
{\it at the Fermi level ($\ep=0$) and at finite temperature}.
Physical properties such as conductance are expressed in terms of
{\it single  particle} states at the Fermi level. It is thus
essential to understand the range of validity of the description
in terms of independent quasiparticles.
 From relation (\ref{taueeT1}), we have  \cite{Eiler}

\be {1 \over \tau_{ee}(T)} = {8 \pi \nu_0^3} \int_0^\infty d
\omega W^2(\omega) {\omega \over \sinh \beta \omega} \ \ .
\label{tauwbeta} \ee
For the diffusion in free space, the matrix element $W^2(\omega)$
is proportional to  $\omega^{d/2}$ (relation \ref{W2q33}), so that

 \be {1 \over   \tau_{ee}(T)}= {\pi d c_d \over 2 \nu_0}
\int_{0}^\infty {d \omega  \over \omega \sinh \beta \omega }
\left( {\omega \over E_c}\right)^{d/2} \ .   \label{tauphiZ3} \ee
Therefore, in three dimensions, we have
 \be {1 \over  \tau_{ee}(T)} = {\sqrt{2} \over 4 \pi^2 \nu_0}
\int_{0}^\infty {d \omega  \over \omega \sinh \beta \omega }
\left( {\omega \over E_c}\right)^{3/2} \simeq { T  \over \nu_0}
\int_{0}^T {d \omega  \over \omega^2} \left( {\omega \over
E_c}\right)^{3/2} \ \ , \label{tauphiT1} \ee
that is

\begin{equation}
\displaystyle  {1 \over \tau_{ee}(T)} \simeq \Delta \left( {T
\over
E_c}\right)^{3/2}\ \ \ \ \ \  \ \ (d=3) \\
\label{tauphi3D} \ee
up to  a numerical factor. Note that the exponent of the power law
is the same as the exponent for the energy dependence of the
lifetime at zero temperature (\ref{AAK}). This result follows at
once if we notice that relevant  processes in the quasiparticle
relaxation described by  $\omega W^2(\omega)$ are those for which
the energy transfer $\omega$ is of order  $T$.

It would be  tempting to generalize this result to any dimension
and to conclude that $1/ \tau_{ee}(T) \propto T^{d/2}$. This is
not correct for $d \leq 2$. In this case, the contribution of e-e
processes with {\it low energy transfer}  $\omega \simeq 0$
dominates and leads to a divergence in the  integral
(\ref{tauphiZ3}). In order to cure this divergence, it is worth
noticing that $\tau_{ee}(T)$ represents precisely the lifetime of
an eigenstate, so that the energy transfer $\omega$ cannot be
defined with an accuracy better than $1/ \tau_{ee}$. Consequently,
there is no energy transfer smaller than $1/\tau_{ee}(T)$, so that
 the integral (\ref{tauphiZ3}) needs to be cut off
self-consistently for $\omega$ smaller than $1/ \tau_{ee}(T)$. For
$d \leq 2$, we thus obtain a self-consistent relation for $
\tau_{ee}$~:
\be {1 \over  \tau_{ee}(T)}\simeq {1 \over \nu_0} \int_{1/
\tau_{ee}}^\infty {d \omega  \over \omega \sinh \beta \omega }
\left( {\omega \over E_c}\right)^{d/2} \simeq {  T   \over \nu_0}
\int_{1/ \tau_{ee}}^T {d \omega  \over \omega^2} \left( {\omega
\over E_c}\right)^{d/2} \label{tauphi1D1}\ee
where   the thermal factor has been replaced by  a cutoff at
$\omega \sim T$. In two dimensions, $1 / \tau_{ee} (T)$ is
proportional to the temperature (within logarithmic corrections)~:
\be {1 \over   \tau_{ee}(T) } \simeq \Delta { T \over E_c} \ln
{E_c \over \Delta}\ \ \ \ \ \ \ \  (d=2) \ \ . \label{tauphi2D}\ee
In one  dimension, and since $T \tau_{ee} \gg 1$, the integral
becomes proportional to  $\sqrt{ \tau_{ee}}$ so that the
self-consistent relation leads to
\begin{equation}
\displaystyle  {1 \over  \tau_{ee}(T)} \simeq \Delta \left({E_c
\over \Delta}\right)^{1/3} \left({ T \over E_c}\right)^{2/3} \ \ \
\ \ \
\ \  (d=1) \ \ .  \\
\label{tauphi1D} \ee

\bigskip

\subsection{Phase coherence} \label{sect.cohee}

 The time (\ref{tauwbeta}) has been defined as the lifetime of a quasiparticle,
 generalizing the notion introduced by Landau to the case of a diffusive
 system in $d$ dimensions. We should now evaluate the phase
 coherence time $\tau_\phi(T)$ which limits coherent effects like
 the weak localization correction (\ref{WL}). This time can be
 interpreted as the lifetime of the Cooperon. Its derivation
 consists in   calculating {\it
directly} the dephasing $\langle e^{i \Phi(t)}\rangle$  resulting
from electron-electron interaction and accumulated between  time
reversed conjugated multiple scattering sequences. To that
purpose,  the interaction between electrons is replaced by an
effective interaction which describes the coupling of a single
electron to the electromagnetic field created by the other
electrons. This calculation \cite{Chakraverty86,Altshuler826} is
not developed here; see \cite{AkkMon} for a detailed derivation.

An alternative and qualitative approach is to consider that  phase
coherence is limited by the lifetime of  quasiparticles.  Since
the multiple scattering trajectories that are paired in the
Cooperon are defined for a given energy state, they cannot
interfere for times larger than $\tau_{ee}(T)$. This results in an
irreversible dephasing between the trajectories and thus a loss of
phase coherence. It is therefore natural to assume that
\be \tau_\phi(T)=\tau_{ee}(T) \  \ . \label{idtemps}\ee
Indeed, the temperature dependences predicted in equations
(\ref{tauphi3D},\ref{tauphi2D},\ref{tauphi1D}) have been confirmed
experimentally, in all dimensions by weak localization
measurements.

It turns out that not only  these two characteristic times
$\tau_{ee}(T)$ and $\tau_\phi(T)$ are equal (within a numerical
factor), but also that the two processes, {\it quasiparticles
relaxation} and {\it phase relaxation}, are very similar. Finally
let us remark that the introduction of the low-energy cutoff in
(\ref{tauphi1D1}) may appear as a handwaving and artiticial way to
handle the low energy divergence.  We have shown recently that the
profound reason for this divergence is that, for $d \leq 2$, {\it
relaxation of quasiparticles} as well as the {\it phase
relaxation} are not exponential \cite{AkkMon1304}.

\medskip

{\bf Acknowledgments - } Many points of view presented in these
lectures have been developed during a long collaboration with Eric
Akkermans, and are detailed in ref. \cite{AkkMon}.


\begin{thebibliography}{99}
\bibitem{exact} Notice that this correct result differs from the
qualitative evaluation (\ref{croix5}) only by a factor $2$.
\bibitem{Cooperon} There is also a contribution in the Cooperon
channel. However, this contribution is corrected by higher order
diagrams and turns out to be small.
\bibitem{lambdas} From our simple picture, the coefficients
$\lambda_\sigma$ and $\lambda_\rho$ are equal. A better
description shows that these parameters are slightly different,
see \cite{AltshulerAronov,AkkMon}.
\bibitem{sumintegral} Alternatively, the sum (\ref{W2q2}) on wavevectors can be replaced by an integral
and we recover (\ref{W2q33}).
\end{thebibliography}
\end{document}